\documentclass[twocolumn,showpacs,showkeys,preprintnumbers]{revtex4-1}
\usepackage{amsmath}
\usepackage{wrapfig}
\usepackage{stackengine,graphicx}
\usepackage{blindtext}
\usepackage{dcolumn}
\usepackage{bm}
\usepackage{amsmath}
\usepackage[position=bottom,caption=false,captionskip=3pt,font=normalsize,
subrefformat=simple,
labelformat=simple]{subfig}
\usepackage{float}

\begin{document}

\title{Magnetic-field-driven topological transitions in
non-centrally-symmetric energy \\
spectrum of 2D electron gas with Rashba-Dresselhaus spin-orbit interaction}
\author{I.V. Kozlov}
\affiliation{B. Verkin Institute for Low Temperature Physics and Engineering of the
National Academy of Sciences of Ukraine, 47 Nauky Ave., Kharkiv 61103,
Ukraine}
\author{Yu.A. Kolesnichenko}
\email{kolesnichenko@ilt.kharkov.ua}
\affiliation{B. Verkin Institute for Low Temperature Physics and Engineering of the
National Academy of Sciences of Ukraine, 47 Nauky Ave., Kharkiv 61103,
Ukraine}

\begin{abstract}
Two dimension electron systems with combined Rashba and Dresselhaus spin
orbit interaction (SOI) having a complicated energy spectrum with a conical
point and four critical points are promising candidates to observe electron
topological transitions. In the present paper we have investigated the
evolution of the electron spectrum and isoenergetic contours under the
influence of parallel magnetic field. General formulas for energies of
critical points for arbitrary values of SOI constants and magnetic field are
found. The existence of critical magnetic fields at which a number of
critical points is changed has been predicted. The magnetic field driving
topological Lifshitz transitions in the geometry of isoenergetic contours
have been studied. Van Hove's singularities in the electron density of
states are calculated. The obtained results can be used for theoretical
investigations of different electron characteristics of such 2D systems.
\end{abstract}

\maketitle



\section{\label{sec:level1} Introduction}

In the fundamental paper \cite{Lifshitz1960} Ilya M. Lifshitz
predicted "electron transitions" - abrupt changes of the Fermi
surface topology under continuous variation of some parameter:
pressure, chemical potential etc. These transitions result in
anomalies in different kinetic and thermodynamic characteristics
of metals that stipulated special attention to their detailed
investigations (see \cite{LAK,Varlamov1989,Blanter1994} for
review). In recent years the interest in Lifshitz transitions get
renewed in view of intensive studies of new electron systems:
graphene, topological insulators, semimetals, superconductors,
Dirac semimetals, and Weyl semimetals, in which different types of
electron topological transitions take place \cite{Volovik2018}.
Low dimensional systems with spin - orbit interaction (SOI)
\cite{Manchon2015,Bercioux2015} are possible candidates to observe
topological transitions in the energy spectrum as well.
Manifestations of topological transitions in a magnetic
susceptibility of
3D semiconductors with SOI had been predicted by Boiko and Rashba \cite%
{Boiko}. Recently the enhanced orbital paramagnetism related to the
topological transition was observed in a layered semiconductor BiTeI when
the Fermi energy $E_{F}$ is near the crossing point of the Rashba spin-split
conduction bands \cite{Schober}\ \

Among a variety of spin - orbit materials the two dimensional (2D) systems
made of zincblende III-V, wurtzite, SiGe semiconductors, semiconductor
quantum wells, etc., occupy a special place possessing combined Rashba -
Dresselhaus (R-D) SOI \cite{Meier2007,Hao2012,Silveira2016,Kepenekian2017}
(see \cite{Ganichev2014} for review). Interplay between two types of SOI
results in anisotropic spin - split Fermi contours which leads to an
anisotropic magnetoresistance \cite{Wang2010}, an enhancement of electron
propagation along a narrow range of real-space angles from an isotropic
source \cite{BermanFlatte2010}, anisotropic Friedel oscillations \cite%
{Badalyan2010,KozKolesnLTP} and so on.

In the presence of SOI a parallel magnetic field $\mathbf{B}$ results not
only in the appearance of a Zeeman energy but also it affects the dispersion
low of charge carriers \cite{Winkler2003}, changing the geometry and
breaking, the central symmetry of 2D Fermi contours \cite{Pershin2004}. It
is a possible way to manipulate with anisotropy of transport characteristics
that seems to be promising for practical applications. The energy spectrum
of 2D electrons with R-D SOI in the in-plane magnetic field $\mathbf{B}$ can
be easily obtained, but to date the information on the evolution of energy
branches with the changes of direction and absolute value of vector $\mathbf{%
B}$ is incomplete and disconnected \cite%
{Tkach2016,ShevchenkoKopeliovich2016,Alekseev2007}. In the recent
paper \cite{Sablikov} an electronic transport in 2D electron gas
subjected to the in-plane magnetic field for the case of Rashba
SOI had been studied theoretically. Singularities of a
conductivity and a spin polarization as functions of the Fermi
level or magnetic field, which occurs when the Fermi level passes
through the Van Hove singularity \cite{vanHove1953}, were
analyzed. It was predicted that the transport anisotropy
dramatically changes near the singularity. Such anisotropy was
reported in an experiment \cite{Wolff}.

In this paper we present a consistent consideration of changes in the energy
spectrum of 2D electrons with R-D SOI under the variations of the parallel
magnetic field. Special attention will be paid to the magnetic-field-induced
2D electron topological transitions. The structure of the paper is as
follows. Section II contains some known information which is the basis of
subsequent investigations. We present the Hamiltonian of the system, its
eigenvalues and eigenfunctions. The energy spectrum in the absence of the
magnetic field is discussed form the point of view of possibilities of
topological transitions. In Sec. III the evolution of energy spectrum for
arbitrary value and direction of vector $\mathbf{B}$ is studied. We predict
the existence of critical fields $B_{c1}$ and $B_{c2}$ at which the number
of critical points (minima and saddle points) of the energy surfaces is
changed. So a possibility to create artificial degenerate critical points of
energy spectrum appears. Limiting cases of weak and strong magnetic field
are considered. In Sec. IV, as examples, we present explicit analytical
results for the energies of critical points, their position in the $\mathbf{k%
}$ - space, and critical values $B_{c1}$ and $B_{c2}$ for the magnetic field
directed along the symmetry axes. Variations of topology of isoenergetic
contours, which are 2D analogue of Fermi surface, under variations of the
magnetic field have been analyzed in Sec. V. In Sec. VI the singular part of
the electron density of states is discussed. The obtained formulas allow to
determine SOI constants from van Hove singularities \cite{vanHove1953}. We
conclude the paper with some final remarks and a summary of main results in
Sec.VII.

\section{\label{sec:level2} Hamiltonian of the system. Energy spectrum at
zero magnetic field}

Our calculations are based on the \ widely used model of the 2D
single-electron Hamiltonian taking into account linear terms of R-D SOI (see, for example, \cite%
{BermanFlatte2010,Badalyan2010,Tkach2016,ShevchenkoKopeliovich2016,Kleinert,Wang,Wen})
\begin{eqnarray}
{{\hat{H}}_{0}} &=&\frac{\widehat{\mathbf{P}}{^{2}}}{2m}{{\sigma }_{0}}+%
\frac{\alpha }{m}({{\sigma }_{x}}\widehat{{P}}{_{y}}-{{\sigma }_{y}}\widehat{%
{P}}{_{x}})  \label{H_0} \\
&&+\frac{\beta }{\hbar }({{\sigma }_{x}}\widehat{{P}}{_{x}}-{{\sigma }_{y}}%
\widehat{{P}}{_{y}})-\frac{{{g}^{\ast }}}{2}{{\mu }_{B}}\mathbf{B\sigma .}
\notag
\end{eqnarray}%
Here $\widehat{\mathbf{P}}=\widehat{\mathbf{p}}+e\mathbf{A}/c$ is the
operator of the generalized momentum, $\widehat{\mathbf{p}}=\hbar \widehat{\mathbf{k}}%
=-i\hbar \mathbf{\nabla =}\left( \widehat{p}_{x},\widehat{p}_{y}\right) $ is
the operator of the in-plane momentum, $\mathbf{A}$ is the vector potential
of the in-plain magnetic field $\mathbf{B}=\left( {{B}_{x}},{{B}_{y}}%
,0\right) $, ${m}$ is an effective electron mass, $\sigma
_{x,y,z}$ are Pauli matrices,\textbf{\ }$\mathbf{\sigma =}\left(
\sigma _{x},\sigma _{y},\sigma
_{z}\right) $ is the Pauli vector,$\ \hat{\sigma}_{0}$ is unit matrix $%
2\times 2$, $\alpha $ and $\beta $ are Rashba ($\alpha $) \cite{Rashba1984}
and Dresselhaus \cite{Dresselhaus1955} $\left( \beta \right) $ constants of
SOI, $\mu _{B}$ is the Bohr magneton, and $g^{\ast }$ is an effective
g-factor of the 2D system. For definiteness we assume $\alpha ,\beta $ to be
nonnegative values and $\alpha \geq \beta $.

In the framework of 2D model the Hamiltonian (\ref{H_0}) does not
depend on a component $P_{z}=p_{z}+eA_{z}/c$ of the generalized
momentum and in the Coulomb gauge, $\mathbf{A}=\left(
0,0,{{B}_{x}}y-{{B}_{y}}x\right) $, $\nabla \cdot \mathbf{A}=0$,
can be rewritten as
\begin{eqnarray}
{{\hat{H}}_{0}} &=&\frac{{{\hbar }^{2}}\left( \hat{k}_{x}^{2}+\hat{k}%
_{y}^{2}\right) }{2m}{{\sigma }_{0}}+\alpha ({{\sigma }_{x}}{{\hat{k}}_{y}}-{%
{\sigma }_{y}}{{\hat{k}}_{x}})  \notag \\
&&+\beta ({{\sigma }_{x}}{{\hat{k}}_{x}}-{{\sigma }_{y}}{{\hat{k}}_{y}})-%
\frac{{{g}^{\ast }}}{2}{{\mu }_{B}}\left( {{B}_{x}}{{\sigma }_{x}}+{{B}_{y}}{%
{\sigma }_{y}}\right) .  \label{spectr}
\end{eqnarray}%
where $\widehat{\mathbf{k}}$ is the wave vector operator.
Neglecting by cubic terms in the Dresselhaus part of Eqs.
(\ref{H_0}), (\ref{spectr}) we assume a narrow quantum well and
small 2D wave vectors $k$ of electrons in a conduction band, $k\ll
\pi /w$, where $w$ is the well width.

The eigenvalues and the eigenfunctions of the Hamiltonian (\ref{spectr}) are
(see, for example, \cite{Tkach2016})
\begin{gather}
\epsilon _{1,2}\left( \mathbf{k}\right) =\frac{\hbar ^{2}k^{2}}{2m}  \notag
\\
\pm \sqrt{(h_{x}+\alpha k_{y}+\beta k_{x})^{2}+(h_{y}-\alpha k_{x}-\beta
k_{y})^{2}},  \label{eigenvalues} \\
{{h}_{x,y}}=-\frac{{{g}^{\ast }}}{2}{{\mu }_{B}}{{B}_{x,y}},  \notag
\end{gather}%
\begin{equation}
\psi _{1,2}\left( \mathbf{r}\right) =\frac{1}{2\pi \sqrt{2}}{{e}^{i\mathbf{kr%
}}}\left( \begin{aligned} & 1 \\ & \pm {{e}^{i\theta }} \end{aligned}\right)
.  \label{eigenfunctions}
\end{equation}

\bigskip The angle $\theta $ defines an average spin direction for two
branches of energy spectrum (\ref{eigenvalues}), $\theta =\theta _{2}=\theta
_{1}+\pi $,

\begin{equation}
\tan \theta =\frac{{{h}_{y}}-\alpha {{k}_{x}}-\beta {{k}_{y}}}{{{h}_{x}}%
+\alpha {{k}_{y}}+\beta {{k}_{x}}},  \label{eigentheta}
\end{equation}%
which depends on the wave vector and the magnetic field.

\begin{figure*}[tbp]
\subfloat {\stackinset{l}{0.31\textwidth}{b}
{0.065\textwidth}{\includegraphics[width=0.16\textwidth]
{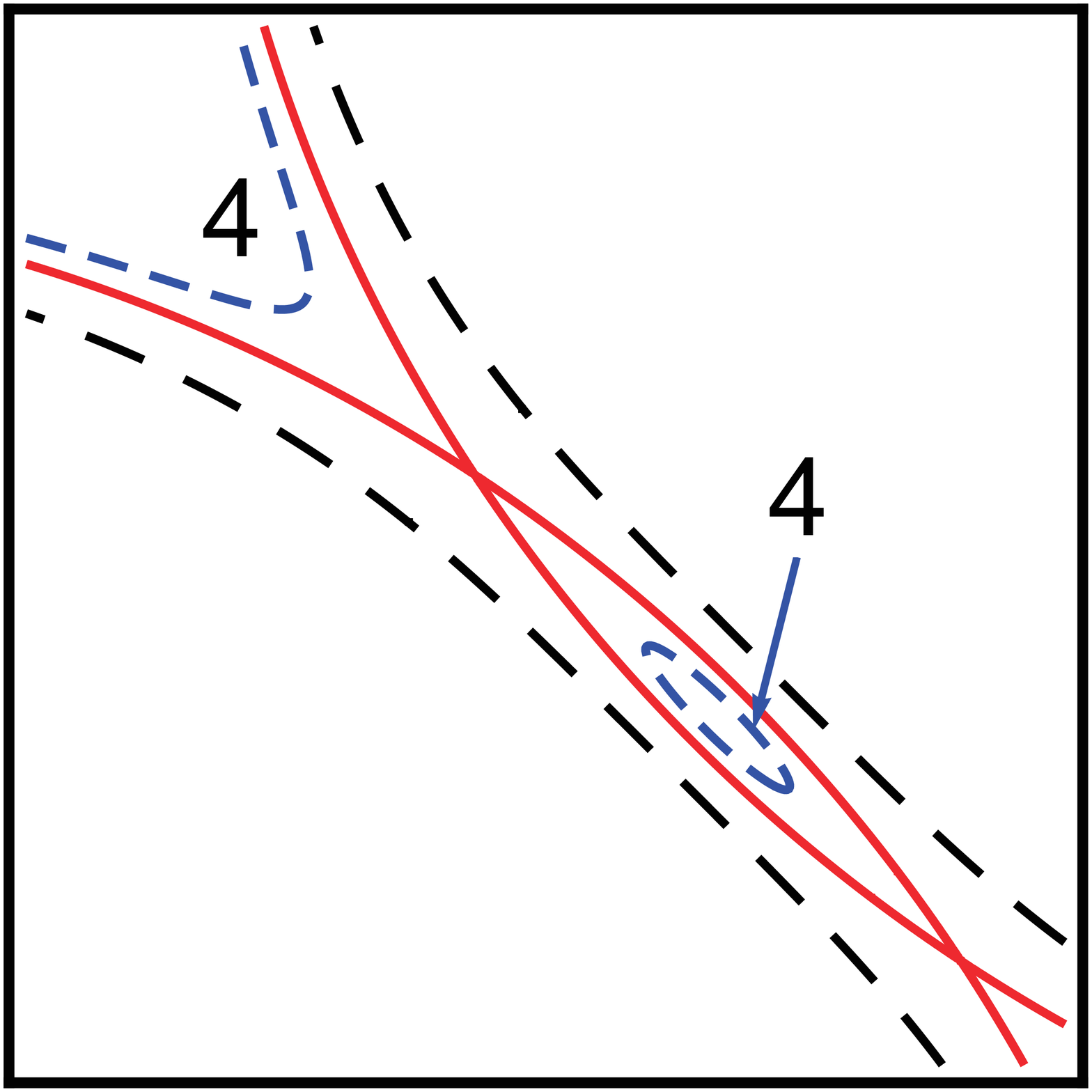}}{\includegraphics[width=0.48\textwidth]{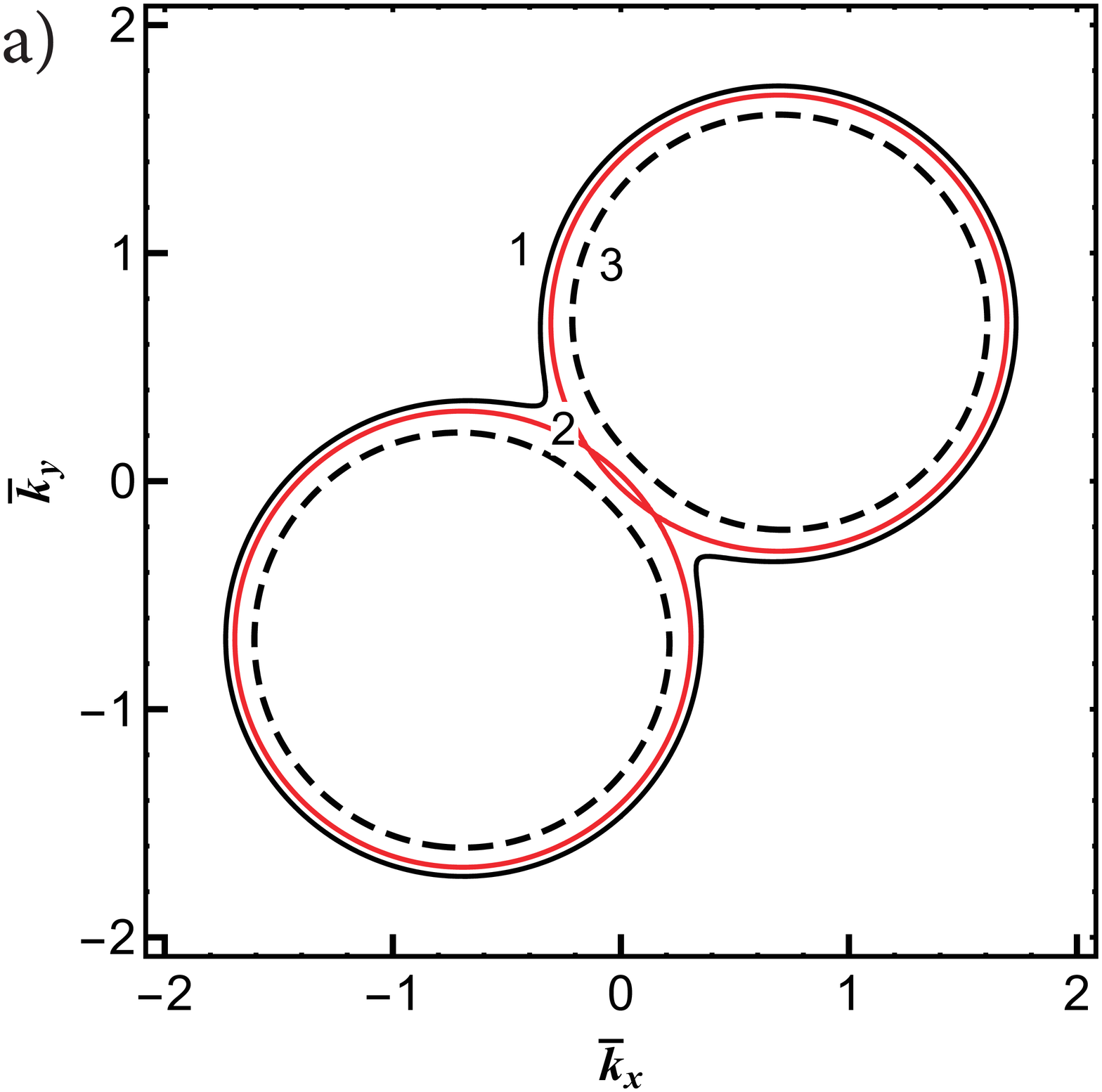}}}
\hfill \subfloat
{\includegraphics[width=0.48\textwidth]{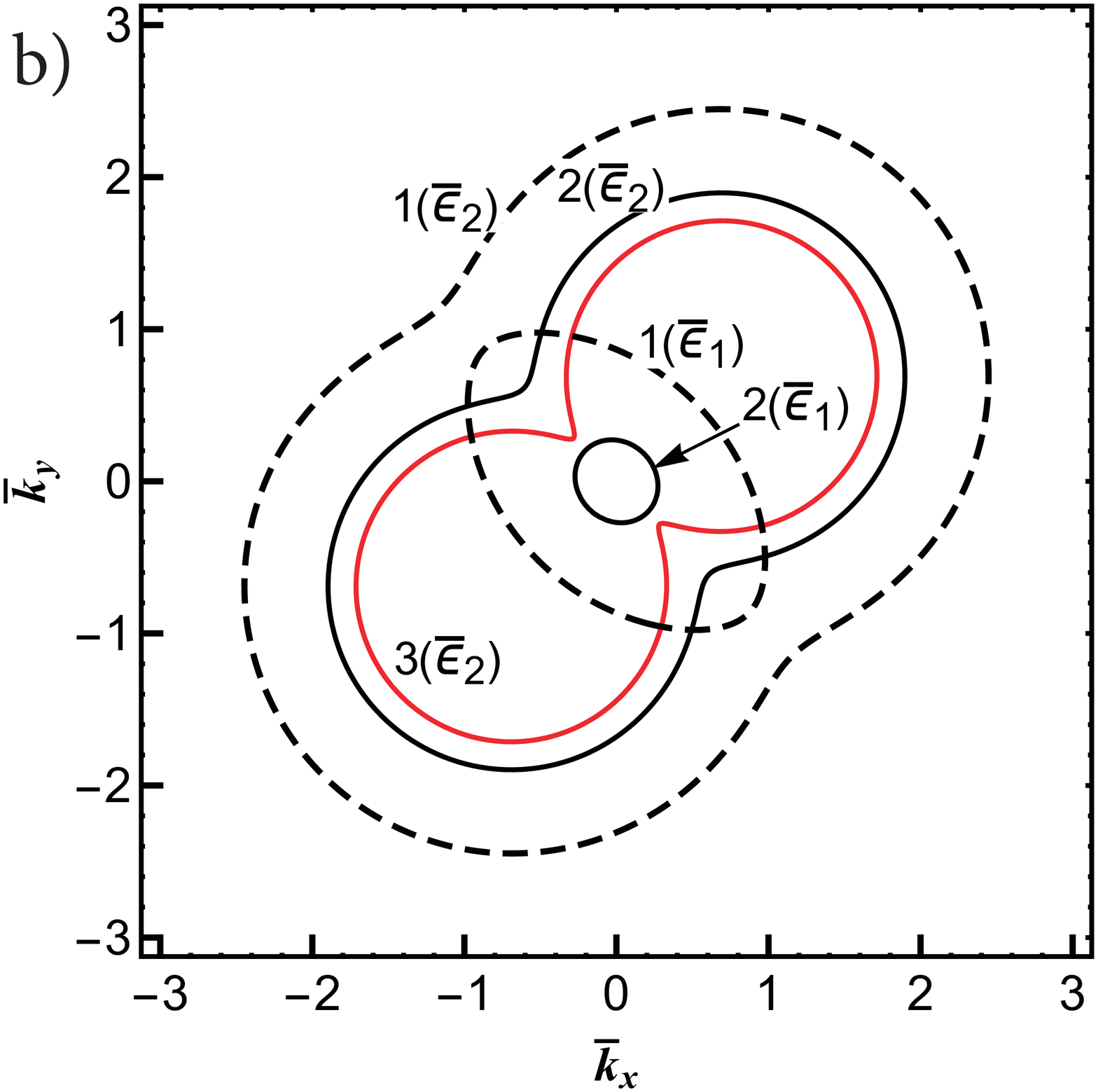}} \quad
\caption{Isoenergetic contours for different energies $E$ at $h=0$, $\bar{%
\protect\epsilon}_{2}^{\min 1,2}=-0.72$, $\bar{\protect\epsilon}%
_{2}^{sad1,2}=-0.02$, $\bar{\protect\alpha}=0.6$, $\bar{\protect\beta}=0.4$.
a) The contours belong to the branch $\protect\epsilon ={{\protect\epsilon }%
_{2}}$ at $E\leq 0$: $E=0$, black solid contour (1); $\bar{E}=\bar{\protect%
\epsilon}_{2}^{sad1,2}=-0.02$, red self-crossed contour (2); $\bar{E}=-0.1$,
dashed contour (3). In the insert: $\bar{E}=-0.01$, blue short-dashed
contour (4). b) The contours ${{\protect\epsilon }_{1,2}}=E$ at $E>0$; $\bar{%
E}=1$, dashed contours (1); $\bar{E}=0.2$, black solid contours (2); $E=0$,
red solid contour (3).}
\label{Fig1}
\end{figure*}
In the absence of magnetic field ${{h}_{x}}={{h}_{y}}=0$ the energy spectrum
(\ref{eigenvalues}) is centrosymmetric ${{\epsilon }_{1,2}}\left( \mathbf{k}%
\right) ={{\epsilon }_{1,2}}\left( -\mathbf{k}\right) $, and has two
symmetry axes ${{k}_{x}}={{k}_{y}}$ and ${{k}_{x}}=-{{k}_{y}}$. At $\alpha
\neq \beta $ energy branches $\epsilon ={{\epsilon }_{1}}\left( \mathbf{k}%
\right) $ and $\epsilon ={{\epsilon }_{2}}\left( \mathbf{k}\right) $
(energies as the function of the wave vector are three dimension surfaces in
$\epsilon ,\ {{k}_{x}},\ {{k}_{y}}$ space) touch each other at the single
point $\mathbf{k}=0$, which is a conical (Dirac) point. In this point the
energy of ${{\epsilon }_{1}}$ branch has a smallest value ${{\epsilon }_{1}}%
\left( 0\right) =0$. The energy branch $\epsilon ={{\epsilon }_{1}}\left(
\mathbf{k}\right) \geq 0$ as the function of wave vector components ${{k}_{x}%
},{{k}_{y}}$ is a convex surface for any values of parameters. The energy
surface corresponding to the second branch $\epsilon ={{\epsilon }_{2}}%
\left( \mathbf{k}\right) $ has two degenerate minima $\epsilon _{2}^{\min
1,2}=-\frac{m}{2{{\hbar }^{2}}}{{\left( \alpha +\beta \right) }^{2}}$ in the
points $k_{x}^{\min 1,2}=k_{y}^{\min 1,2}=\mp \frac{m}{\sqrt{2}{{\hbar }^{2}}%
}\left( \alpha +\beta \right) $, two saddle points $%
k_{x}^{sad1,2}=-k_{y}^{sad1,2}=\pm \frac{m}{\sqrt{2}{{\hbar }^{2}}}\left(
\alpha -\beta \right) $ corresponding to the energy $\epsilon _{2}^{sad1,2}=-%
\frac{m}{2{{\hbar }^{2}}}{{\left( \alpha -\beta \right) }^{2}}$, and the
conical point ${{\epsilon }_{2}}\left( 0\right) =0$ at $\mathbf{k}=0$.

The isoenergetic contours of constant energy ${{\epsilon }_{1,2}}\left(
\mathbf{k}\right) =E$ (2D contours of a constant energy $\epsilon =E$ in the
${{k}_{x}},\ {{k}_{y}}$ plane) are a 2D analogue of 3D Fermi surface
pockets. At positive energies $E>0$ the spectrum has two spin - split
contours (see Fig. 1b). The larger contour ($1\left( {{\epsilon }_{2}}%
\right) $, $2\left( {{\epsilon }_{2}}\right) $ in Fig. 1b) belongs to the
branch $\epsilon ={{\epsilon }_{2}}$. The smaller contour ($1\left( {{%
\epsilon }_{1}}\right) $, $2\left( {{\epsilon }_{1}}\right) $ in Fig. 1b) of
the branch $\epsilon ={{\epsilon }_{1}}$ is always situated inside the
larger one. For ${{\epsilon }_{2}}\left( \mathbf{k}\right) =E<0$ the
isoenergetic contour become not simply connected: In the range $\epsilon
_{2}^{sad1,2}<E<0$ inside the larger isoenergetic contour the contour of
smaller radius is situated (contour 4 in the insert in Fig. 1a). The
electron velocity on this contour is directed along an inner normal, i.e.
this contour can be interpreted as a "hole" one. At $E=\epsilon
_{2}^{sad1,2} $the contours becomes self-crossed. In the range $\epsilon
_{2}^{\min 1,2}<E<\epsilon _{2}^{sad1,2}$ this contour splits into two parts
which do not span the point $\mathbf{k}=0$. The contours of the branch $%
\epsilon ={{\epsilon }_{2}}\left( \mathbf{k}\right) $ is non-convex in the
energy interval \cite{KozKolesnLTP}
\begin{equation}
\epsilon _{2}^{sad1,2}\leq E<-\frac{m{{\left( \alpha +\beta \right) }^{2}}%
\left( {{\alpha }^{2}}-6\alpha \beta +{{\beta }^{2}}\right) }{2{{\hbar }^{2}}%
{{\left( \alpha -\beta \right) }^{2}}}.  \label{uslEsedl}
\end{equation}%
In the special case, $\alpha =\beta $, two energy spectrum branches contact
along the parabola ${{\epsilon }_{1}}={{\epsilon }_{2}}=\frac{{{\hbar }^{2}}{%
{k}^{2}}}{2m}$ in the plane crossing the symmetry axis ${{k}_{x}}=-{{k}_{y}}$
.

Plotting different dependencies in this paper we use for numerical
computations the dimensionless values
\begin{eqnarray}
&&\bar{\alpha}=\frac{m\alpha }{{{\hbar }^{2}}{{k}_{0}}},\quad \bar{\beta}=%
\frac{m\beta }{{{\hbar }^{2}}{{k}_{0}}},\quad \bar{h}=\frac{h}{{{\epsilon }%
_{0}}},  \notag \\
&&\bar{k}=\frac{k}{{{k}_{0}}},\quad \bar{\epsilon}=\frac{\epsilon }{{{%
\epsilon }_{0}}},\quad {{k}_{0}}=\sqrt{2m{{\epsilon }_{0}}}/\hbar ,
\label{defk0}
\end{eqnarray}%
where $\epsilon _{0}>0$ is a constant of energy dimension, for example, an
absolute value of Fermi energy.

Figure 1 demonstrates the full set of Lifshitz transitions under changes of
the energy: the appearance (or disappearance) of the new detached region at $%
E=0$ (Fig. 1b), disruption (or formation) of the contour "neck" (Fig. 1a)
and the appearance of the critical self-crossing contour at $\epsilon
_{2}^{sad1,2}=E$ (contour 2 in Fig. 1a).

The concentration of 2D electron gas created in heterostructures, and hence
the Fermi energy, can be controlled by means of a gate electrode. Why a
magnetic field is needed? In the system with spin - orbit interaction an
electric field perpendicular to the plane of 2D electrons not only shifts
the Fermi level but also changes the Rashba SOI constant \cite%
{Nitta1997,Sato2001,Beukman2017}. That may make the interpretation of
experimental results ambiguous. The parallel magnetic field plays the role
of independent parameter which can tune critical points of the energy
spectrum to the Fermi energy. In the next sections we consider possibilities
to drive by characteristics of energy spectrum of 2D electron gas with R-D
SOI by means of in-plain magnetic field .

\section{\label{sec:level3} Arbitrary magnetic field direction. General
relations}

In the parallel magnetic field and $\alpha \neq \beta $ the point of energy
branch contact moves from the point $\mathbf{k}=0$ to the point $\mathbf{k}={%
{\mathbf{k}}_{0}}$ whose coordinates of which must be found from the
condition
\begin{equation}
\sqrt{{{\left( {{h}_{x}}+\alpha {{k}_{y0}}+\beta {{k}_{x0}}\right) }^{2}}+{{%
\left( {{h}_{y}}-\alpha {{k}_{x0}}-\beta {{k}_{y0}}\right) }^{2}}}=0.
\label{uslRoot}
\end{equation}%
It is easy to see that Eq. (\ref{uslRoot}) is equivalent to the system of
linear inhomogeneous equations
\begin{equation}
\alpha {{k}_{y0}}+\beta {{k}_{x0}}=-{{h}_{x}};\quad \alpha {{k}_{x0}}+\beta {%
{k}_{y0}}={{h}_{y}},  \label{h0}
\end{equation}%
and we have
\begin{eqnarray}
&&{{k}_{x0}}=h\frac{\alpha \sin {{\varphi }_{h}}+\beta \cos {{\varphi }_{h}}%
}{{{\alpha }^{2}}-{{\beta }^{2}}},  \notag \\
&&{{k}_{y0}}=-h\frac{\alpha \cos {{\varphi }_{h}}+\beta \sin {{\varphi }_{h}}%
}{{{\alpha }^{2}}-{{\beta }^{2}}},  \label{k0}
\end{eqnarray}%
where angle ${{\varphi }_{h}}$ defines the magnetic field direction $\mathbf{%
h}=h\left( \cos {{\varphi }_{h}},\sin {{\varphi }_{h}},0\right) $. The
energy value corresponding to the point $\mathbf{k}={{\mathbf{k}}_{0}}%
=\left( {{k}_{x0}},{{k}_{y0}}\right) $ (\ref{k0}) is given by
\begin{equation}
{{\epsilon }_{1}}\left( {{\mathbf{k}}_{0}}\right) ={{\epsilon }_{2}}\left( {{%
\mathbf{k}}_{0}}\right) ={{E}_{0}}={{h}^{2}}{{\hbar }^{2}}\frac{{{\alpha }%
^{2}}+{{\beta }^{2}}+2\alpha \beta \sin 2{{\varphi }_{h}}}{2m{{({{\alpha }%
^{2}}-{{\beta }^{2}})}^{2}}}.  \label{E0}
\end{equation}%
If the SOI constants are equal, $\alpha =\beta $, the equations (\ref{h0})
have nonzero solutions only if ${{h}_{x}}=-{{h}_{y}}$. In this case the
branches contact along the parabola
\begin{eqnarray}
&&{{\epsilon }_{cont}}\left( {{k}_{y1}},h\right) =\frac{{{\hbar }^{2}}%
k_{y1}^{2}}{2m}+\frac{{{\hbar }^{2}}{{h}^{2}}}{8m{{\alpha }^{2}}},  \notag \\
&&{{k}_{x1}}=\frac{{{k}_{x}}+{{k}_{y}}}{\sqrt{2}}=\frac{h}{2\alpha },\quad {{%
k}_{y1}}=\frac{{{k}_{x}}-{{k}_{y}}}{\sqrt{2}}.  \label{Econt}
\end{eqnarray}%
For any other directions of the vector $\mathbf{h}$ the branches do not have
common points for $\alpha =\beta $.

For further analysis of the energy spectrum at $\alpha \neq \beta $ it is
useful to introduce polar coordinates $\tilde{k}>0,$ and $f$ with the center
in the point ${{\mathbf{k}}_{0}}$ (\ref{k0}):
\begin{equation}
{{k}_{x}}=k_{x0}+\tilde{k}\cos f,\quad {{k}_{y}}=k_{y0}+\tilde{k}\sin f.
\label{ktilde}
\end{equation}

Note that new coordinates only shift the energy spectrum in \textbf{k} -
space and they don't change differential characteristics of $\epsilon ={{%
\epsilon }_{1,2}}\left( \mathbf{k}\right) $ surfaces.

In coordinates $\tilde{k},f$ (\ref{ktilde}) the energies ${{\epsilon }_{1,2}}
$ take the simple form
\begin{equation}
{{\epsilon }_{1,2}}\left( \tilde{k},\tilde{f}\right) =\frac{{{\hbar }^{2}}{{{%
\tilde{k}}}^{2}}}{2m}-\frac{{{\hbar }^{2}}\tilde{k}}{m}{{\lambda }_{1,2}}%
\left( f\right) +{{E}_{0}},  \label{E12}
\end{equation}%
where
\begin{eqnarray}
{{\lambda }^{\left( 1,2\right) }}\left( f\right) &=&h\frac{\alpha \sin
\left( f-{{\varphi }_{h}}\right) -\beta \cos \left( f+{{\varphi }_{h}}%
\right) }{{{\alpha }^{2}}-{{\beta }^{2}}}  \notag \\
&&\mp \frac{m}{{{\hbar }^{2}}}\sqrt{{{\alpha }^{2}}+{{\beta }^{2}}+2\alpha
\beta \sin (2f)},  \label{lambda}
\end{eqnarray}%
\begin{equation}
{{\lambda }^{\left( 2\right) }}\left( f\right) =-{{\lambda }^{\left(
1\right) }}\left( f+\pi \right) .  \label{lambdasim}
\end{equation}%
Substituting Eqs. (\ref{ktilde}) into formula (\ref{eigentheta}) one finds
that spin directions are antisymmetric, $\theta \left( f+\pi \right) =\theta
\left( f\right) +\pi $, with respect to the point ${{\mathbf{k}}_{0}}=\left(
{{k}_{x0}},{{k}_{y0}}\right) $.

The sign of the Gauss curvature ${{K}^{\left( 1,2\right) }}\left( \tilde{k}%
,f\right) $ of energy surfaces $\epsilon ={{\epsilon }_{1,2}}\left( \tilde{k}%
,\tilde{f}\right) $ is defined by the sign of the determinant $\det \left(
H\right) $ of the Hessian matrix
\begin{equation}
H=\left( \begin{aligned} && \frac{{{\partial }^{2}}{{\epsilon
}_{1,2}}}{\partial k_{x}^{2}} && \frac{{{\partial }^{2}}{{\epsilon
}_{1,2}}}{\partial k_{x}^{{}}\partial {{k}_{y}}} && \\ && \frac{{{\partial
}^{2}}{{\epsilon }_{1,2}}}{\partial {{k}_{y}}\partial {{k}_{x}}} &&
\frac{{{\partial }^{2}}{{\epsilon }_{1,2}}}{\partial k_{y}^{2}} &&
\end{aligned}\right) .  \label{Hmatrix}
\end{equation}%
In coordinates (\ref{ktilde}) one finds
\begin{equation}
\det \left( H\right) =\frac{{{\hbar }^{4}}}{{{m}^{2}}\tilde{k}}\left[ \tilde{%
k}-{{{\ddot{\lambda}}}^{\left( 1,2\right) }}(f)-{{\lambda }^{\left(
1,2\right) }}(f)\right] .  \label{detH}
\end{equation}%
Here and in all formulas below the points above functions denote the
derivative with respect to the angle $\,f$ . From Eq. (\ref{lambda}) one can
see that the sum of ${{\lambda }^{\left( 1,2\right) }}$ and its second
derivatives ${{\ddot{\lambda}}^{\left( 1,2\right) }}$ do not depend on
magnetic field and has the definite sign
\begin{equation}
{{\ddot{\lambda}}^{\left( 1,2\right) }}(f)+{{\lambda }^{\left( 1,2\right) }}%
(f)=\mp \frac{m}{{{\hbar }^{2}}}\frac{{{\left( {{\alpha }^{2}}-{{\beta }^{2}}%
\right) }^{2}}}{{{\left( {{a}^{2}}+2\alpha \beta \sin (2f)+{{\beta }^{2}}%
\right) }^{3/2}}}.  \label{lambdaddot}
\end{equation}%
From Eqs. (\ref{detH}), (\ref{lambdaddot}) it follows that ${{K}^{\left(
1\right) }}\left( \tilde{k},f\right) >0$ for any values of parameters, i.e.
the surface $\epsilon ={{\epsilon }_{1}}\left( \tilde{k},\tilde{f}\right) $
is the convex one.

\begin{figure*}[tbp]
\subfloat {\includegraphics[width=0.48\textwidth]{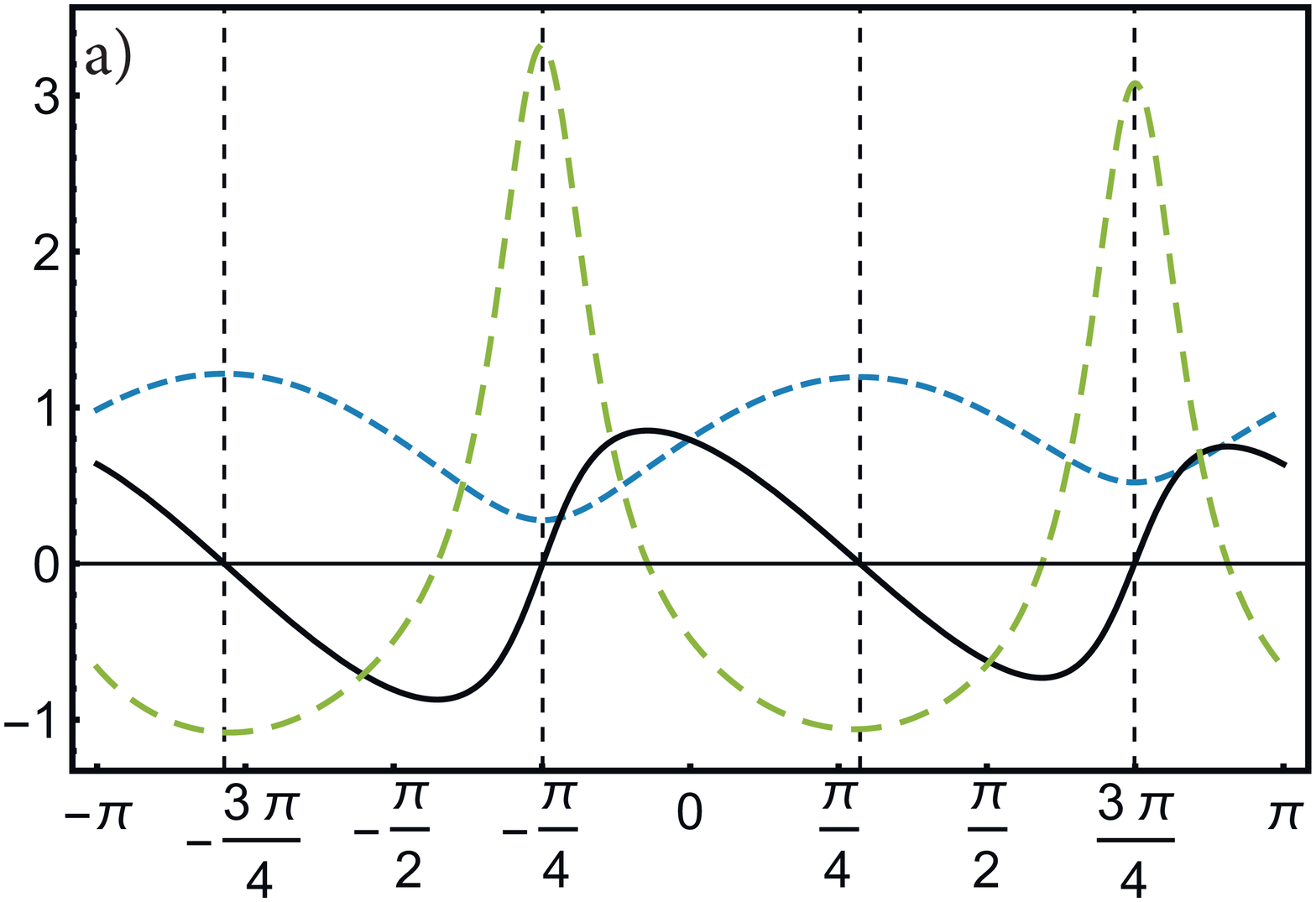}} \quad
\subfloat
{\includegraphics[width=0.48\textwidth]{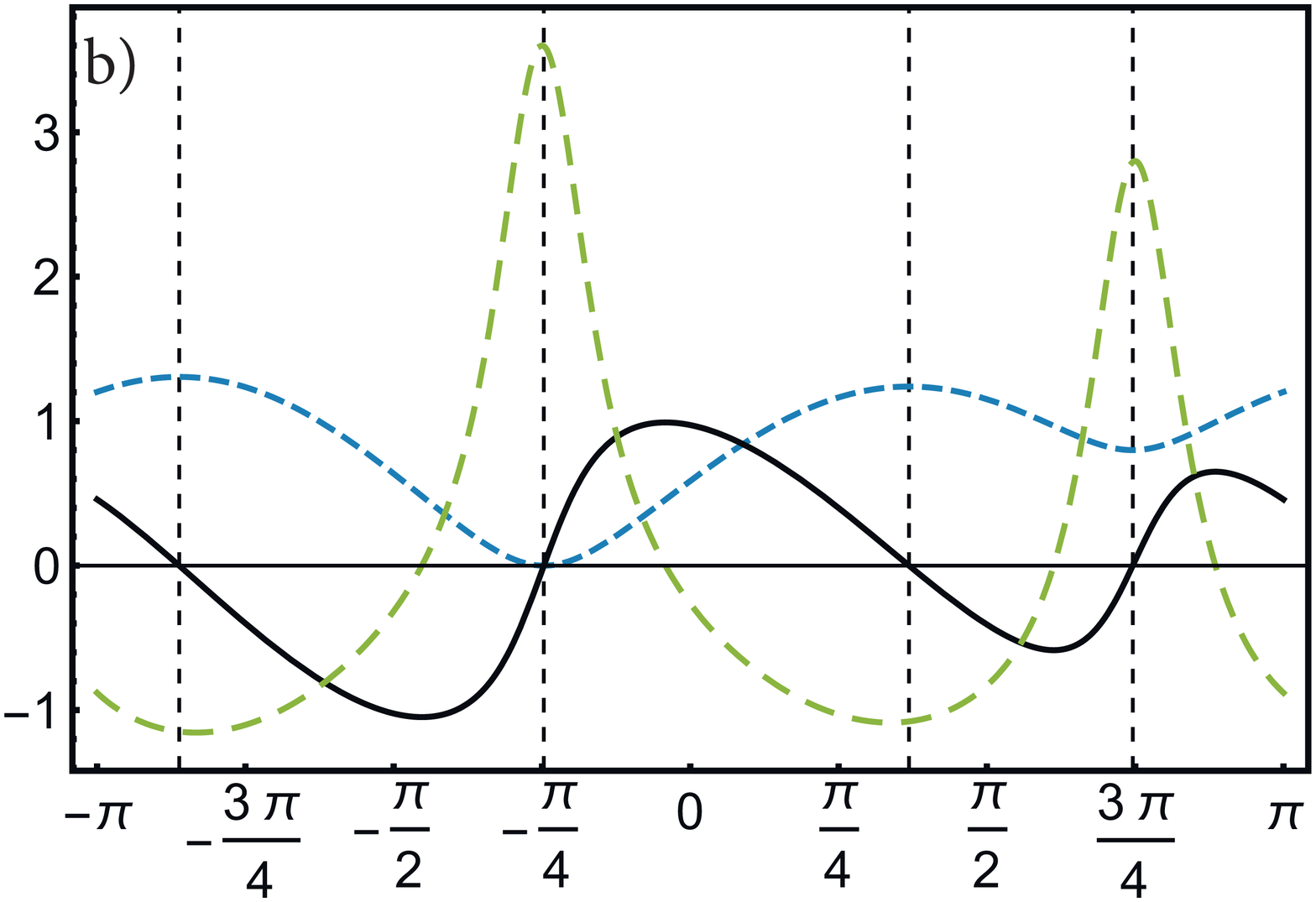}} \quad \subfloat
{\includegraphics[width=0.48\textwidth]{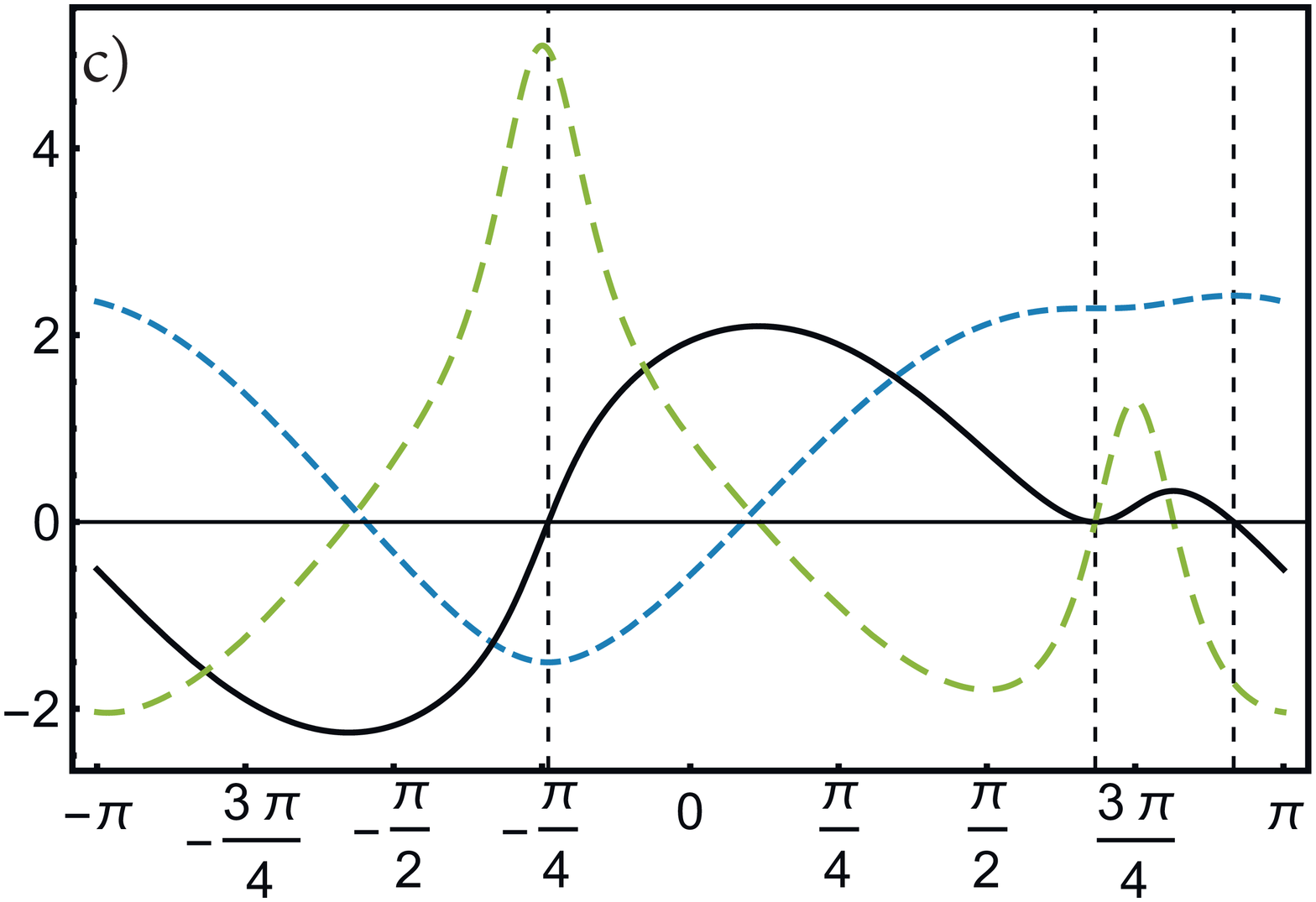}} \quad \subfloat
{\includegraphics[width=0.48\textwidth]{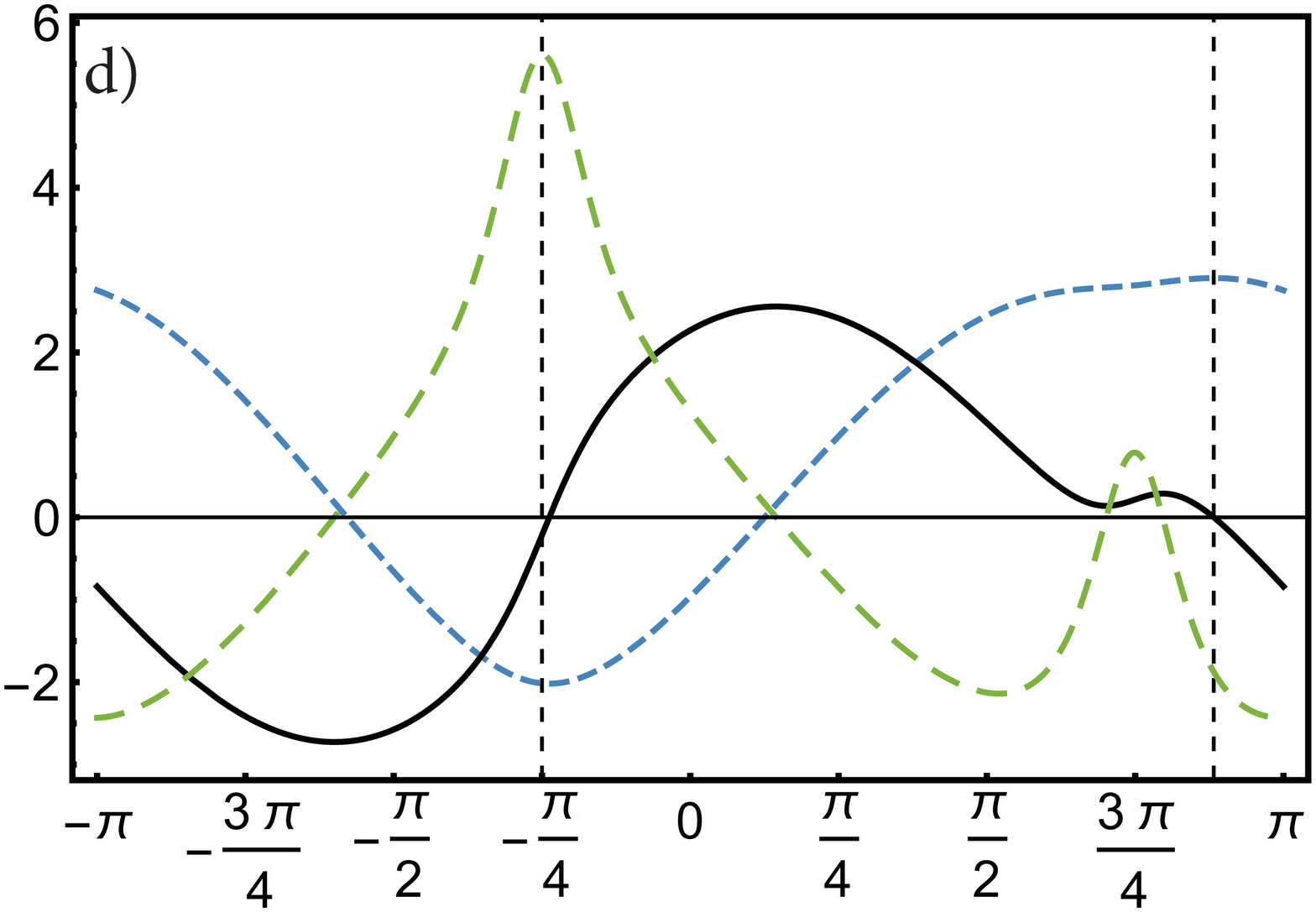}} \quad
\caption{The dependencies of the function ${{\protect\lambda }_{2}}\left(
f\right) $ (\protect\ref{lambda}) (short dashed lines), its first ${{\dot{%
\protect\lambda}}_{2}}\left( f\right) $ (solid lines) and second ${{\ddot{%
\protect\lambda}}_{2}}\left( f\right) $ (long dashed lines) derivatives on
the angle $f\in \left[ -\protect\pi ,\protect\pi \right] $ for different
values of magnetic field $h$: a) $\bar{h}=0.05<{{\bar{h}}_{c2}}=0.1655$; b) $%
h={{h}_{c2}}$; c) $\bar{h}={{\bar{h}}_{c1}}=0.7867$; d) $\bar{h}=1>{{\bar{h}}%
_{c1}}$. Vertical dashed lines show angles $f=f_{\protect\nu }^{\left(
2\right) }$ corresponding ${{\dot{\protect\lambda}}_{2}}\left( f_{\protect%
\nu }^{\left( 2\right) }\right) =0$. For SOI constants and magnetic field
direction we used the values $\bar{\protect\alpha}=0.8$, $\bar{\protect\beta}%
=0.4$, ${{\protect\varphi }_{h}}=\protect\pi /3$.}
\label{fig2}
\end{figure*}

Critical points $\left( {{{\tilde{k}}}_{\nu }},{{f}_{\nu }}\right) $ of the
energy spectrum should be found from the system of equations
\begin{equation}
\frac{\partial {{\epsilon }_{1,2}}}{\partial \tilde{k}}=\frac{{{\hbar }^{2}}%
}{m}\left( \tilde{k}-{{\lambda }^{\left( 1,2\right) }}\left( f\right)
\right) =0,\,\tilde{k}\geq 0,  \label{dEdk}
\end{equation}%
\begin{equation}
\frac{\partial {{\epsilon }_{1,2}}}{\partial f}=-\frac{{{\hbar }^{2}}}{m}%
\tilde{k}{{\dot{\lambda}}^{\left( 1,2\right) }}\left( f\right) =0,
\label{dEdf}
\end{equation}%
from which we give
\begin{equation}
\tilde{k}_{\nu }^{\left( 1,2\right) }={{\lambda }^{\left( 1,2\right) }}%
\left( f_{\nu }^{\left( 1,2\right) }\right) \;(a),\;{{\dot{\lambda}}^{\left(
1,2\right) }}\left( f_{\nu }^{\left( 1,2\right) }\right) =0\;(b);
\label{kcr}
\end{equation}%
where index $\nu $ numerates the roots of Eq.(\ref{dEdf}). According to the
definition the variable $\tilde{k}$ is an absolute value of electron wave
vector in coordinates (\ref{ktilde}). Only solutions for which $\tilde{k}%
_{\nu }^{\left( 1,2\right) }={{\lambda }^{\left( 1,2\right) }}\left( f_{\nu
}^{\left( 1,2\right) }\right) >0$ have the physical meaning.

The obvious equality Eq.(\ref{lambdasim}) gives the relations between
solutions (\ref{kcr}):
\begin{eqnarray}
&&f_{\nu }^{\left( 1\right) }=f_{\nu }^{\left( 2\right) }+\pi ,\quad \tilde{k%
}\left( f_{\nu }^{\left( 1\right) }\right) =-\tilde{k}\left( f_{\nu
}^{\left( 2\right) }\right) ,  \notag \\
&&{{\ddot{\lambda}}^{\left( 1\right) }}\left( f_{\nu }^{\left( 1\right)
}\right) =-{{\ddot{\lambda}}^{\left( 2\right) }}\left( f_{\nu }^{\left(
2\right) }\right) .  \label{sim}
\end{eqnarray}%
The determinant $\det \left( H\right) $ (\ref{detH}) in critical points
reads
\begin{equation}
\det \left( H\left( {{{\tilde{k}}}_{\nu }},{{f}_{\nu }}\right) \right) ={{%
\left. -\frac{{{\hbar }^{4}}}{{{m}^{2}}\tilde{k}}{{{\ddot{\lambda}}}^{\left(
1,2\right) }}\left( f\right) \right\vert }_{\tilde{k}={{{\tilde{k}}}_{\nu }}%
,f=f_{\nu }^{\left( 1,2\right) }}},\,\tilde{k}>0.  \label{detHcr}
\end{equation}

If ${{\ddot{\lambda}}^{\left( 1,2\right) }}\neq 0$, the critical point is
nondegenerate. From Eq. (\ref{detHcr}) we conclude that the negative second
derivative ${{\ddot{\lambda}}^{\left( 1,2\right) }}\left( f_{\nu }^{\left(
1,2\right) }\right) <0$ corresponds to energy minima ${{\epsilon }_{1,2}}%
\left( {{{\tilde{k}}}_{\nu }},f_{\nu }^{\left( 1,2\right) }\right) =\epsilon
_{1,2}^{\min }$ and for saddle points of non-convex surface ${{\epsilon }_{2}%
}\left( {{{\tilde{k}}}_{\nu }},f_{\nu }^{\left( 1,2\right) }\right)
=\epsilon _{2}^{sad}$ the second derivative is positive ${{\ddot{\lambda}}%
^{\left( 2\right) }}\left( f_{\nu }^{\left( 1,2\right) }\right) >0$. As it
easy to see from the Eqs. (\ref{E12}),(\ref{kcr}) the energies in critical
points are written as
\begin{equation}
\epsilon _{1,2}^{crit}={{\epsilon }_{1,2}}\left( {{{\tilde{k}}}_{\nu }}%
,f_{\nu }^{\left( 1,2\right) }\right) ={{E}_{0}}-\frac{{{\hbar }^{2}}}{2m}{{%
\left( {{\lambda }^{\left( 1,2\right) }}\left( f_{\nu }^{\left( 1,2\right)
}\right) \right) }^{2}},  \label{Ecrit}
\end{equation}%
i.e. all critical points are situated below energy level $E={{E}_{0}}$. So,
the evolution of either energy branch of 2D electrons with R-D SOI under
influence of parallel magnetic field is completely described by means of the
single function $\lambda ^{\left( 1,2\right) }\left( f\right) $ (\ref{lambda}%
).

For arbitrary magnetic field the equation (\ref{kcr}b) can be transformed to
quartic equation for $\cos \left( 2f_{\nu }^{\left( 1,2\right) }\right) $,
the exact solutions of which are well known. Unfortunately they are so
lengthy that not suitable to any analytical calculation. Nevertheless for
numerical computations the solution of equation (\ref{kcr}b) presents no
problems.

The limiting cases of weak and strong magnetic fields can be analyzed by
means of expansions of exact eigenenergies (\ref{eigenvalues}). For the weak
magnetic field the power expansion of energy ${{\epsilon }_{2}}$ (\ref%
{eigenvalues}) on $\,h$ gives energies of critical points and their
positions. As a results of direct calculations one obtains the following
expressions for two minima
\begin{eqnarray}
&&\epsilon _{2}^{\min 1,2}\simeq -\frac{m}{2{{\hbar }^{2}}}{{(\alpha +\beta )%
}^{2}}\mp h\sin \left( {{\varphi }_{h}}-\frac{\pi }{4}\right) ,  \notag \\
&&h\ll \frac{m}{{{\hbar }^{2}}}{{\left( \alpha +\beta \right) }^{2}},
\label{Emin}
\end{eqnarray}%
\begin{equation}
k_{x}^{\min 1,2}=k_{y}^{\min 1,2}\simeq \mp \frac{m}{{{\hbar }^{2}}}\frac{%
\alpha +\beta }{\sqrt{2}}-\frac{\left( \alpha -\beta \right) h\sin \left( {{%
\varphi }_{h}}+\frac{\pi }{4}\right) }{\sqrt{2}{{\left( \alpha +\beta
\right) }^{2}}},  \label{kmin}
\end{equation}%
and two saddle points
\begin{eqnarray}
&&\epsilon _{2}^{sad1,2}\simeq -\frac{m}{2{{\hbar }^{2}}}{{(\alpha -\beta )}%
^{2}}\pm h\sin \left( {{\varphi }_{h}}+\frac{\pi }{4}\right) ,  \notag \\
&&h\ll \frac{m}{{{\hbar }^{2}}}{{\left( \alpha -\beta \right) }^{2}},
\label{Esad}
\end{eqnarray}%
\begin{equation}
k_{x}^{sad1,2}=-k_{y}^{sad1,2}\simeq \pm \frac{\alpha -\beta }{\sqrt{2}}%
\frac{m}{{{\hbar }^{2}}}-\frac{h(\alpha +\beta )\sin \left( {{\varphi }_{h}}-%
\frac{\pi }{4}\right) }{\sqrt{2}{{(\alpha -\beta )}^{2}}}.  \label{ksad}
\end{equation}%
The branch ${{\epsilon }_{1}}$ in this case hasn't extremum. The energy ${{%
\epsilon }_{1}}$ reaches the least value ${{\epsilon }_{1}}\left( {{\mathbf{k%
}}_{0}}\right) ={{E}_{0}}$ (\ref{E0}) in the point of branch contact ${{%
\mathbf{k}}_{0}}=\left( {{k}_{x0}},{{k}_{y0}}\right) $ (\ref{k0}).

In the strong magnetic field $h\gg \frac{m}{{{\hbar }^{2}}}\left( {{\alpha }%
^{2}}+{{\beta }^{2}}-2\alpha \beta \sin 2{{\varphi }_{h}}\right) $ the power
series of ${{\epsilon }_{1,2}}$ (\ref{eigenvalues}) on $1/h$ gives the
energy minima $\epsilon _{1,2}^{\min }$ of both branches
\begin{equation}
\epsilon _{1,2}^{\min }\simeq \pm h-\frac{m}{2{{\hbar }^{2}}}\left( {{\alpha
}^{2}}+{{\beta }^{2}}-2\alpha \beta \sin 2{{\varphi }_{h}}\right) ,
\label{EminS}
\end{equation}%
\begin{eqnarray}
&&k_{x1,2}^{\min }\simeq \pm \frac{m}{{{\hbar }^{2}}}\left( \alpha \sin {{%
\varphi }_{h}}-\beta \cos {{\varphi }_{h}}\right) ,  \notag \\
&&k_{y1,2}^{\min }\simeq \pm \frac{m}{{{\hbar }^{2}}}\left( \beta \sin {{%
\varphi }_{h}}-\alpha \cos {{\varphi }_{h}}\right) .  \label{kminS}
\end{eqnarray}%
So, with the increasing of the magnetic field the energy spectrum evolves
from the energy branch ${{\epsilon }_{2}}$ having four critical points and
the branch ${{\epsilon }_{1}}$ without critical points to the spectrum every
branch of which has single critical (minimum) point. How such evolution
occurs? For arbitrary magnetic field direction some general conclusions can
be made of the basis of the properties of functions ${{\lambda }^{\left(
1,2\right) }}\left( f\right) $ (\ref{lambda}) and its derivatives ${{\dot{%
\lambda}}^{\left( 1,2\right) }}\left( f\right) $, ${{\ddot{\lambda}}^{\left(
1,2\right) }}\left( f\right) $.

As it has been concluded above the branch $\epsilon ={{\epsilon }_{1}}$ is
convex and hasn't saddle points. It is clear from Eq. (\ref{lambda}) at weak
magnetic field ($h\rightarrow 0$) ${{\lambda }^{\left( 1\right) }}\left(
f\right) <0$ for any angle $f$, i.e. for energy branch $\epsilon ={{\epsilon
}_{1}}\left( \mathbf{k}\right) $ the Eq. (\ref{dEdk}) hasn't positive
solutions $\tilde{k}>0$. The critical value $h={{h}_{c2}}$ can be derived by
means of the equation ${{\lambda }^{\left( 1\right) }}\left( h={{h}_{c2}}%
;f\right) =0$ from which we find
\begin{equation}
{{h}_{c2}}=\frac{{{\left( {{\alpha }^{2}}-{{\beta }^{2}}\right) }^{2}}}{%
\sqrt{{{\alpha }^{4}}+6{{\alpha }^{2}}{{\beta }^{2}}+{{\beta }^{4}}+4\alpha
\beta \left( {{\alpha }^{2}}+{{\beta }^{2}}\right) \text{sin2}{{\varphi }_{h}%
}}}.  \label{hc2}
\end{equation}%
Substituting ${{\lambda }^{\left( 1\right) }}\left( f\right) $ (\ref{lambda}%
) in Eq. (\ref{Ecrit}) and taking into account the positiveness of the
function ${{\lambda }^{\left( 1\right) }}\left( f_{\nu }^{\left( 1\right)
}\right) >0$ at $h>{{h}_{c2}}$, it is easy to show that $0\leq \epsilon
_{1}^{\min }\leq {{E}_{0}}$ for any values of parameters. As it follows from
Eq. (\ref{sim}) the appearance of minima of the branch ${{\epsilon }_{1}}$
is accompanied by the disappearance of saddle point of the branch ${{%
\epsilon }_{2}}$.

The energy surface $\epsilon ={{\epsilon }_{2}}$ has regions of negative
Gauss curvature. The derivative ${{\dot{\lambda}}^{\left( 2\right) }}\left(
f\right) $ is the sum the oscillatory functions with periods $2\pi $ and $%
\pi $. Depending on the magnetic field value it has two zeros at $h>{{h}_{c1}%
}$ and four zeros at $h<{{h}_{c1}}$ in the range $\left[ -\pi ,\pi \right] $%
. Numbers of zeros ${{\dot{\lambda}}^{\left( 2\right) }}\left( f\right) $
having different sign of second derivative ${{\ddot{\lambda}}^{\left(
2\right) }}\left( f\right) $ are equal. The critical value ${{h}_{c1}}$ can
be found from the condition of the coalescence of two zeros of ${{\dot{%
\lambda}}^{\left( 2\right) }}\left( h,f\right) $ with different signs of
second derivative ${{\ddot{\lambda}}^{\left( 2\right) }}\left( h;f\right) $
which in this point vanishes, i.e. one should search two unknown quantities $%
{{h}_{c1}},\ {{f}_{c}}$ from two equations
\begin{equation}
{{\dot{\lambda}}^{\left( 2\right) }}\left( {{h}_{c1}},{{f}_{c}}\right)
=0;\quad {{\ddot{\lambda}}^{\left( 2\right) }}\left( {{h}_{c1}},{{f}_{c}}%
\right) =0.  \label{lambdaziro}
\end{equation}%
We couldn't find the analytical solution of this system for arbitrary
magnetic field orientation. The critical fields ${{h}_{c1}}$ in an explicit
form for special directions of vector $\mathbf{h}$ are found in the next
section. Note that the sign of the difference ${{h}_{c2}}-{{h}_{c1}}$ is not
fixed for given $\alpha ,\ \beta $ and depends on the magnetic field
direction.

The evolution of the branch $\epsilon ={{\epsilon }_{2}}\left( \mathbf{k}%
\right) $ can be understood from Fig. 2. In the magnetic field $h<\min
\left( {{h}_{c1}},{{h}_{c2}}\right) $, $\tilde{k}_{\nu }^{\left( 2\right) }={%
{\lambda }^{\left( 2\right) }}\left( f_{\nu }^{\left( 2\right) }\right) >0$,
and there are four critical points ${{\dot{\lambda}}^{\left( 2\right) }}%
\left( f_{\nu }^{\left( 2\right) }\right) =0$ - two minima $\epsilon
_{2}^{\min 1,2}$ $\left( {{{\ddot{\lambda}}}^{\left( 1,2\right) }}\left( {{f}%
_{\nu }}\right) <0\right) $ and two saddle points $\epsilon _{2}^{sad1,2}$ $%
\left( {{{\ddot{\lambda}}}^{\left( 1,2\right) }}\left( {{f}_{\nu }}\right)
>0\right) $ (Fig. 2a). At $h={{h}_{c2}}$ one of the saddle points coincides
with the point of branch contact, $\tilde{k}_{\nu }^{\left( 2\right) }={{%
\lambda }^{\left( 2\right) }}\left( f_{\nu }^{\left( 2\right) }\right) =0$,
and "disappear" (Fig.2b). In the magnetic field $h={{h}_{c1}}$ the minimum
and saddle points of energy surface $\epsilon ={{\epsilon }_{2}}$
"annihilate" (Fig.2c) and in larger fields $h>\max \left( {{h}_{c1}},{{h}%
_{c2}}\right) $ the branch $\epsilon ={{\epsilon }_{2}}\left( \mathbf{k}%
\right) $ has one absolute minimum (Fig. 2d).

In next paragraph we consider some cases when exact formulas become an
elementary and give clear illustrations of general conclusion of this
paragraph.

\section{\label{sec:level4} Magnetic field along symmetry axis}

For the direction of the magnetic field along one of the symmetry axis the
energy spectrum preserves the symmetry with respect to other axis. This
circumstance essentially simplify the solution of equations obtained above.

\textbf{a) } \textsl{Magnetic field directed along ${{k}_{x}}=-{{k}_{y}}$
axis }

Let us consider the magnetic field direction ${{\varphi }_{h}}=3\pi /4$. In
this case the equation (\ref{dEdf}), from which the angles ${{f}_{\nu }}$ ($%
\nu =1,2,3,4$) corresponding to zeros of derivative ${{\lambda }_{1,2}}$ can
be found. The Eq. (\ref{kcr}b) has four solutions in the interval $\left[
-\pi ,\pi \right] $. Two solutions do not depend on magnetic field and SOI
constants
\begin{equation}
f_{1}^{\left( 1,2\right) }=-\frac{3\pi }{4},\quad f_{2}^{\left( 1,2\right) }=%
\frac{\pi }{4},  \label{fsim}
\end{equation}%
and two solutions exist in the finite interval of value $h\in \left[ 0,{{h}%
_{c1}}\right] $
\begin{eqnarray}
&&f_{3}^{\left( 1,2\right) }\left( h\right) =-\frac{\pi }{4}\mp \arcsin
\left( \eta \right) ,  \notag \\
&&f_{4}^{\left( 1,2\right) }\left( h\right) =\frac{3\pi }{4}\pm \arcsin
\left( \eta \right) ,  \label{f34}
\end{eqnarray}
where
\begin{eqnarray}
&&\eta \left( \alpha ,\beta ,h\right) =\frac{h(\alpha -\beta )}{2\sqrt{%
\alpha \beta \left( h_{c1}{{h}_{c2}}-{{h}^{2}}\right) }},  \notag \\
&&\eta \leq 1\Leftrightarrow h\leq {{h}_{c1}}.  \label{f34eta}
\end{eqnarray}%
The functions ${%
{\lambda }^{\left( 2\right) }}\left( f_{\nu }^{\left( 1,2\right)
}\right)$ and ${{\ddot{\lambda}}^{\left( 2\right) }}\left( f_{\nu
}^{\left( 1,2\right) }\right)$ which define the energy of critical
points and their character are
presented in Appendix. The critical magnetic fields in the case under consideration are%
\begin{equation}
{{h}_{c1}}=\frac{4m}{{{\hbar }^{2}}}\alpha \beta ;\quad {{h}_{c2}}=\frac{m}{{%
{\hbar }^{2}}}{{\left( \alpha +\beta \right) }^{2}}.  \label{f34h}
\end{equation}%

\begin{figure*}[tbp]
\subfloat {\includegraphics[width=0.48\textwidth]{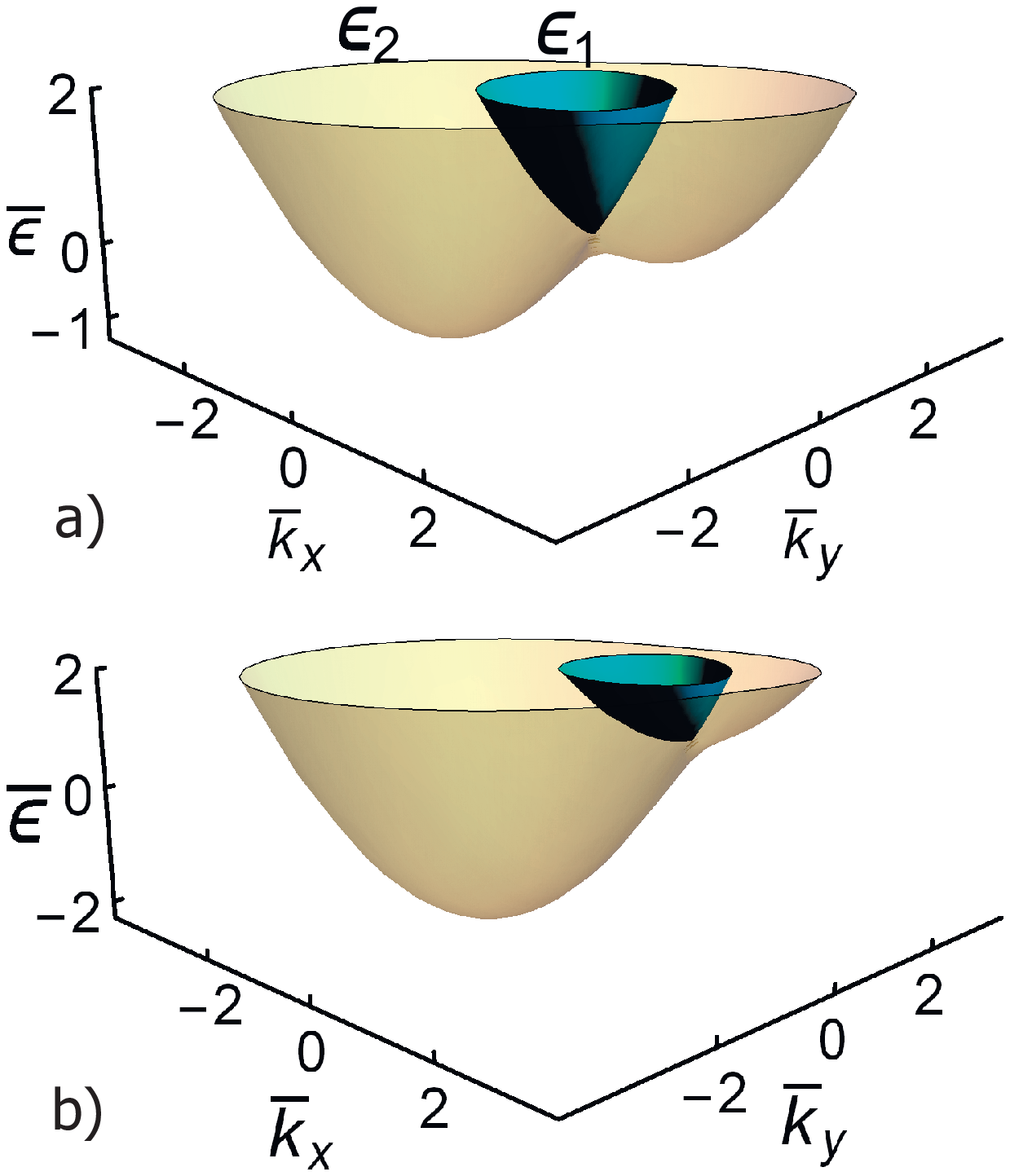}} \quad
\subfloat
{\includegraphics[width=0.48\textwidth]{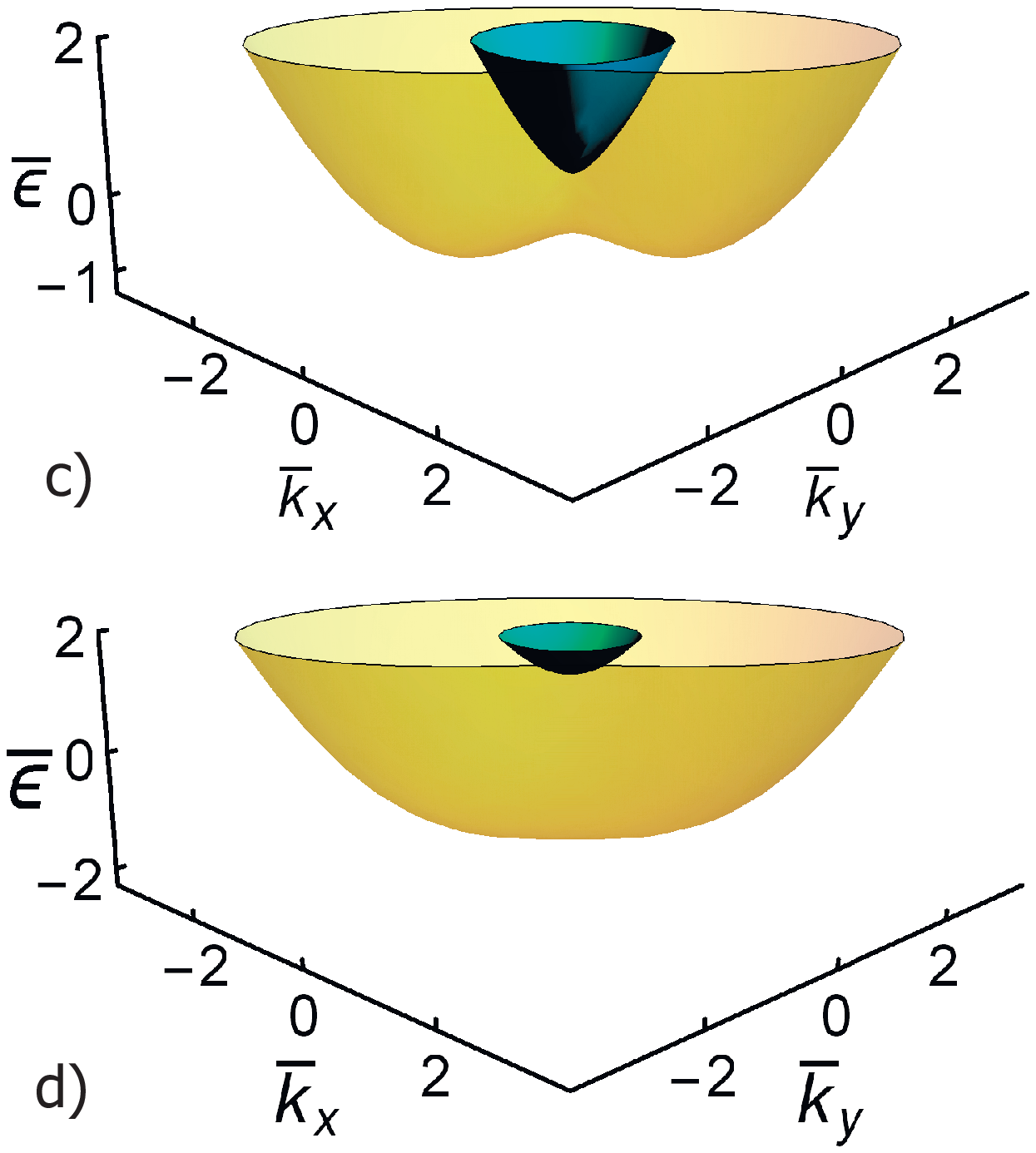}} \quad
\caption{{Energy spectrum (\protect\ref{eigenvalues}) for the magnetic field
directed along ${{k}_{x}}=-{{k}_{y}}$ axis, ${{\protect\varphi }_{h}}=3%
\protect\pi /4$, (a,b) and along ${{k}_{x}}={{k}_{y}}$ axis, ${{\protect%
\varphi }_{h}}=\protect\pi /4$, (c,d). a,c) $\bar{h}=0.5<{{\bar{h}}_{c1}}%
=1.28,\ {{\bar{h}}_{c2}}=1.44$; b,d) $\bar{h}=1.5>{{\bar{h}}_{c1}},\ {{\bar{h%
}}_{c2}}$. }For SOI constants we used the values $\bar{\protect\alpha}=0.8$,
$\bar{\protect\beta}=0.4$. }
\label{fig3}
\end{figure*}

In accordance with Eqs. (\ref{Ecrit}),(\ref{fsimlambda}) for any value of
magnetic field the energy of the branch $\epsilon ={{\epsilon }_{2}}$ has
the minimum $\epsilon _{2}^{\min 1}$
\begin{eqnarray}
&&\epsilon _{2}^{\min 1}\left( h\right) ={{\epsilon }^{crit1}}\left(
h\right) ,\qquad f_{1}^{\left( 2\right) }=-\frac{3\pi }{4};  \notag \\
&&\tilde{k}_{1}^{\left( 2\right) }\left( h\right) =\frac{m}{{{\hbar }^{2}}}%
\left( \alpha +\beta \right) +\frac{h}{\alpha +\beta }.  \label{Emin1}
\end{eqnarray}%
With an increase in magnetic field the minimum $\epsilon _{2}^{\min 1}$
moves down. The second minimum of this branch $\epsilon _{2}^{\min 2}$
\begin{eqnarray}
&&\epsilon _{2}^{\min 2}\left( h\right) ={{\epsilon }^{crit2}}\left(
h\right) ;\qquad f_{2}^{\left( 2\right) }=\frac{\pi }{4},  \notag \\
&&\tilde{k}_{2}^{\left( 2\right) }\left( h\right) =\frac{m}{{{\hbar }^{2}}}%
\left( \alpha +\beta \right) -\frac{h}{\alpha +\beta },  \label{Emin2}
\end{eqnarray}%
exists in the field interval $0\leq h<{{h}_{c1}}$. In this interval the
branch ${{\epsilon }_{2}}$ has two saddle points with equal energies
\begin{eqnarray}
&&\epsilon _{2}^{sad3,4}\left( h\right) ={{\epsilon }^{crit3,4}}\left(
h\right) ,\qquad f=f_{3,4}^{\left( 2\right) },  \notag \\
&&\tilde{k}_{3,4}^{\left( 2\right) }\left( h\right) =\frac{m}{{{h}^{2}}}%
(\alpha -\beta )\sqrt{1-\frac{{{h}^{2}}}{{{h}_{c1}}{{h}_{c2}}}}.
\label{Emin34}
\end{eqnarray}%
In the field $h\rightarrow {{h}_{c1}}$ the minimum $\epsilon _{2}^{\min 2}$
transforms to the saddle point $\epsilon _{2}^{\min 2}\rightarrow \epsilon
_{2}^{sad}={{\epsilon }^{crit2}}$ (the second derivative ${{\ddot{\lambda}}%
^{\left( 2\right) }}$ (\ref{fsimdot}) changes the sign at $h={{h}_{c1}}$)
which blends with two saddle points $\epsilon _{2}^{sad3,4}$ (\ref{Emin34}),
i.e. the critical point becomes degenerate
\begin{gather}
{{\epsilon }^{\min 2}}\left( {{h}_{c1}}\right) =\epsilon _{2}^{sad}\left( {{h%
}_{c1}}\right) =\epsilon _{2}^{sad3,4}\left( {{h}_{c1}}\right)  \notag \\
=-\frac{m}{2{{\hbar }^{2}}}{{\left( \alpha ^{2}-6\alpha \beta +\beta
^{2}\right) }^{2}},  \notag \\
\tilde{k}_{2}^{\left( 2\right) }\left( {{h}_{c1}}\right) =\tilde{k}%
_{3,4}^{\left( 2\right) }\left( {{h}_{c1}}\right) =\frac{m\left( \alpha
-\beta \right) ^{2}}{\hbar ^{2}(\alpha +\beta )},  \label{Ecrit1} \\
f_{2}^{\left( 2\right) }\left( {{h}_{c1}}\right) =f_{3,4}^{\left( 2\right)
}\left( {{h}_{c1}}\right) =\frac{\pi }{4}.  \notag
\end{gather}%
In larger fields ${{h}_{c1}}<h<{{h}_{c2}}$ the saddle point $\epsilon
_{2}^{sad}$ exists. At $h\rightarrow {{h}_{c2}}$ its energy $\epsilon
_{2}^{sad}\rightarrow {{E}_{0}}$ \ and this saddle point disappears $\ $in
the field $h={{h}_{c2}}$ ( ${{\lambda }^{\left( 2\right) }}$ (\ref%
{fsimlambda}) becomes negative at $h>{{h}_{c2}}$)%
\begin{eqnarray}
{{\epsilon }^{crit2}}\left( {{h}_{c2}}\right) &=&\epsilon _{2}^{sad}\left( {{%
h}_{c2}}\right) ={{E}_{0}}\left( {{h}_{c2}}\right)  \notag \\
&=&\frac{m}{2{{\hbar }^{2}}}{{\left( \alpha +\beta \right) }^{2}},
\label{Ecrit2} \\
\tilde{k}_{2}^{\left( 2\right) }\left( {{h}_{c2}}\right) &=&0;\quad
f_{2}^{\left( 2\right) }=\frac{\pi }{4}.  \notag
\end{eqnarray}

In the magnetic fields $h>{{h}_{c2}}$ the function ${{\lambda }^{\left(
1\right) }}\left( \pi /4\right) $ (\ref{fsimlambda}) becomes positive and
first energy branch ${{\epsilon }_{1}}$ acquires the critical point
(minimum) $\epsilon _{1}^{\min }={{\epsilon }^{crit2}}$
\begin{eqnarray}
&&\epsilon _{1}^{\min }\left( h\right) ={{\epsilon }^{crit2}},\quad
f_{2}^{\left( 1\right) }=\frac{\pi }{4},  \notag \\
&&\tilde{k}_{2}^{\left( 1\right) }=-\frac{m}{{{\hbar }^{2}}}\left( \alpha
+\beta \right) +\frac{h}{\alpha +\beta }.  \label{EminHigh}
\end{eqnarray}%
The minimum $\epsilon _{1}^{\min }$ moves up with increase in the magnetic
field. The discussed evolution of energy spectrum is illustrated in Fig. 3.

\textbf{b)} \textsl{Magnetic field directed along ${{k}_{x}}={{k}_{y}}$ axis}

In this case the solutions of the Eq. (\ref{kcr}b) take the form (we choose $%
{{\varphi }_{h}}=\pi /4$)
\begin{eqnarray}
&&f_{1}^{\left( 1,2\right) }=\frac{3\pi }{4},\;f_{2}^{\left( 1,2\right) }=-%
\frac{\pi }{4},\;f_{3}^{\left( 1,2\right) }\left( h\right) =\frac{\pi }{4}%
\mp \arcsin \left( \eta \right) ,  \notag \\
&&f_{4}^{\left( 1,2\right) }\left( h\right) =\frac{5\pi }{4}\pm \arcsin
\left( \eta \right) ,  \label{f1234b}
\end{eqnarray}%
where
\begin{eqnarray}
&&\eta \left( \alpha ,\beta ,h\right) =\frac{h(\alpha +\beta )}{2\sqrt{%
\alpha \beta \left( {{h}_{c1}}{{h}_{c2}}+{{h}^{2}}\right) }},  \notag \\
&&\quad \eta \leq 1\Leftrightarrow h\leq {{h}_{c1}}.  \label{f34etab}
\end{eqnarray}%
The critical magnetic fields are%
\begin{equation}
{{h}_{c1}}=\frac{4m}{{{\hbar }^{2}}}\alpha \beta ,\quad {{h}_{c2}}=\frac{m}{{%
{\hbar }^{2}}}{{\left( \alpha -\beta \right) }^{2}}.  \label{f34hb}
\end{equation}%

As it follows from formulas (\ref{lambda12b})-(\ref{lambda34dotb}) in Appendix, at $h<{{h%
}_{c2}}$ the branch $\epsilon ={{\epsilon }_{2}}$ has two minima $\epsilon
_{2}^{\min 1,2}$ with equal energies
\begin{eqnarray}
&&\epsilon _{2}^{\min 1,2}\left( h\right) =-\frac{m}{2{{\hbar }^{2}}}{{%
\left( \alpha +\beta \right) }^{2}}-\frac{{{h}^{2}}{{\hbar }^{2}}}{8m\alpha
\beta },\quad h\leq {{h}_{c1}},  \notag \\
&&k_{3,4}^{\left( 2\right) }=\frac{m}{{{h}^{2}}}(\alpha +\beta )\sqrt{1+%
\frac{{{h}^{2}}}{{{h}_{c1}}{{h}_{c2}}}};\quad f=f_{3,4}^{\left( 2\right) },
\label{Emin12b}
\end{eqnarray}%
and two saddle points $\epsilon _{2}^{sad1,2}$ (see Fig. 3a)
\begin{eqnarray}
&&\epsilon _{2}^{sad1,2}\left( h\right) =-\frac{m}{2{{\hbar }^{2}}}{{\left(
\alpha -\beta \right) }^{2}}\mp h,  \notag \\
&&\tilde{k}_{1,2}^{\left( 2\right) }\left( f_{1,2}^{\left( 2\right) }\right)
=\frac{m}{{{\hbar }^{2}}}\left( \alpha -\beta \right) \mp \frac{h}{\alpha
-\beta },  \notag \\
&&f_{1}^{\left( 2\right) }=-\frac{\pi }{4},\quad f_{2}^{\left( 2\right) }=%
\frac{3\pi }{4}.  \label{Esad12b}
\end{eqnarray}%
In magnetic field $h={{h}_{c1}}$ the both minima $\epsilon _{2}^{\min 1,2}$
and the saddle point $\epsilon _{2}^{sad1}$ transform into one degenerate
critical point
\begin{equation}
\epsilon _{2}^{\min 1,2}\left( {{h}_{c1}}\right) =\epsilon _{2}^{sad2}\left(
{{h}_{c1}}\right) =-\frac{m}{2{{\hbar }^{2}}}\left( {{\alpha }^{2}}+6\alpha
\beta +{{\beta }^{2}}\right) ,  \label{Emin12c1b}
\end{equation}%
\begin{equation}
f_{3}^{\left( 4\right) }\left( {{h}_{c1}}\right) =f_{4}^{\left( 2\right)
}\left( {{h}_{c1}}\right) =f_{2}^{\left( 2\right) }=\frac{3\pi }{4},
\label{f234c1b}
\end{equation}%
\begin{eqnarray}
&&\widetilde{{k}}{_{2}^{\left( 2\right) }}\left( {{h}_{c1}}\right) =%
\widetilde{{k}}{_{3,4}^{\left( 2\right) }}\left( {{h}_{c1}}\right) =\frac{m}{%
{{\hbar }^{2}}}\frac{{{\left( \alpha +\beta \right) }^{2}}}{\alpha -\beta };
\notag \\
&&{{\ddot{\lambda}}^{\left( 2\right) }}\left( {{h}_{c1}},\frac{3\pi }{4}%
\right) ={{\ddot{\lambda}}^{\left( 2\right) }}\left( {{h}_{c1}}%
,f_{3,4}^{\left( 2\right) }\right) =0,  \label{lambdac1b}
\end{eqnarray}%
and for larger fields $h>{{h}_{c1}}$ one minimum
\begin{eqnarray}
&&\epsilon _{2}^{\min }\left( h\right) =-\frac{m}{2{{\hbar }^{2}}}{{\left(
\alpha -\beta \right) }^{2}}-h,  \notag \\
&&\tilde{k}_{1}^{\left( 2\right) }\left( f_{1}^{\left( 2\right) }\right) =%
\frac{m}{{{\hbar }^{2}}}\left( \alpha -\beta \right) -\frac{h}{\alpha -\beta
},\;f_{1}^{\left( 2\right) }=-\frac{\pi }{4},  \label{Eminllb}
\end{eqnarray}%
and one saddle point remains
\begin{eqnarray}
&&\epsilon _{2}^{sad}\left( h\right) =-\frac{m}{2{{\hbar }^{2}}}{{\left(
\alpha -\beta \right) }^{2}}+h,  \notag \\
&&\tilde{k}_{2}^{\left( 2\right) }\left( f_{2}^{\left( 2\right) }\right) =%
\frac{m}{{{\hbar }^{2}}}\left( \alpha -\beta \right) +\frac{h}{\alpha -\beta
},\;f_{2}^{\left( 2\right) }=\frac{3\pi }{4},  \label{Esadllb}
\end{eqnarray}%
which exists till $h<{{h}_{c2}}$.

The first branch $\epsilon ={{\epsilon }_{1}}$ reaches the smallest value
\begin{equation}
{{E}_{0}}=\frac{{{h}^{2}}{{\hbar }^{2}}}{2m{{\left( \alpha -\beta \right) }%
^{2}}},  \label{E0b}
\end{equation}%
at the weak magnetic field $h<{{h}_{c2}}$. If $h>{{h}_{c2}}$, this branch
has the minimum (see Fig. 3b)
\begin{equation}
\epsilon _{1}^{\min }\left( h\right) =-\frac{m}{2{{\hbar }^{2}}}{{\left(
\alpha -\beta \right) }^{2}}+h,  \label{Eminggb}
\end{equation}%
\begin{equation}
\tilde{k}_{1}^{\left( 1\right) }=-\frac{m}{{{\hbar }^{2}}}\left( \alpha
-\beta \right) +\frac{h}{\alpha -\beta },\quad f_{1}^{\left( 1\right) }=%
\frac{3\pi }{4}.  \label{kminggb}
\end{equation}%
The results of this paragraph and Fig.3 illustrate a quite different
evolution of energy spectrum for the same values of SOI constants but
different magnetic field directions.

\section{\label{sec:level5} Isoenergetic contours}

The dispersion relation of 2D electron gas can be characterized by
isoenergetic contours ${{\epsilon }_{1,2}}=E=const$. According to the theory
of electron topological transitions \cite{LAK,Varlamov1989} when the energy
level $E$ crosses the energy of the critical point $\epsilon _{1,2}^{crit}$
isoenergetic contours change their topology. By analogy with 3D case we name
2D contours at $E=\epsilon _{1,2}^{crit}$ as critical contours. The critical
contours always have the point $\left( {{{\tilde{k}}}_{\nu }},{{f}_{\nu }}%
\right) $ (\ref{kcr}) in which the electron velocity $\mathbf{v}=\frac{%
\partial {{\epsilon }_{1,2}}}{\hbar \partial \mathbf{k}}=0$,
\begin{eqnarray}
&&\left\vert \mathbf{v}\right\vert =\sqrt{{{\left( \frac{\partial {{\epsilon
}_{1,2}}}{\hbar \partial {{k}_{x}}}\right) }^{2}}+{{\left( \frac{\partial {{%
\epsilon }_{1,2}}}{\hbar \partial {{k}_{y}}}\right) }^{2}}}  \notag \\
&=&\frac{\hbar }{m}\sqrt{\left[ {{(\tilde{k}-{{\lambda }^{\left( 1,2\right) }%
}(f))}^{2}}+{{\left( {{{\dot{\lambda}}}^{\left( 1,2\right) }}(f)\right) }^{2}%
}\right] }.  \label{vabs}
\end{eqnarray}%
In this regard a self-crossing contour is not a critical one because $%
\mathbf{v}\neq 0$ at the cross point. However, one should remember that the
cross point is a particular point in a vicinity of which the electron
dispersion is linear in the wave vector components. For the minimum point $%
\epsilon _{1,2}^{crit}=\epsilon _{1,2}^{\min }$ the contour is absent for
energy $E<\epsilon _{1,2}^{\min }$, while for saddle points $\epsilon
_{2}^{crit}=\epsilon _{2}^{sad}$ contours exist both at $E<\epsilon
_{2}^{sad}$ and $E>\epsilon _{2}^{sad}$.

\begin{figure}
\centering\includegraphics[width=0.4\textwidth]{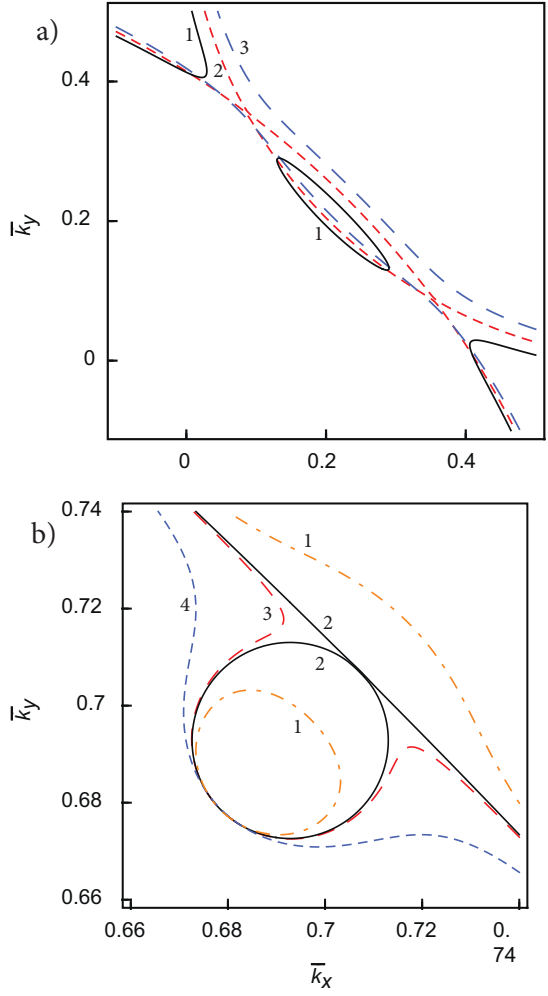}
\caption{Fine structure of isoenergetic contours of the branch
$\protect\epsilon ={{\protect\epsilon }_{2}}$ for different
magnetic fields $\bar{h}\le
{{\bar{h}}_{c1}}=0.96<{{\bar{h}}_{c2}}=1$. a) The energy $E\le
{{E}_{0}}$ equals to the energy of the saddle point
$\bar{E}=\bar{\protect\epsilon }_{2}^{sad3,4}\left( \bar{h}=0.3
\right)=0.026875 $ (\protect\ref{Emin34}): $\bar{h}=0.29$, black
solid line (1); $\bar{h}=0.3$, red short-dashed line (2);
$\bar{h}=0.31$, blue long-dashed line (3).  b) The energy
$\bar{E}<{{E}_{0}}\left( {{h}_{c1}} \right)=0.4608$ equals the
energy of the saddle point $\bar{E}=\bar{\protect\epsilon
}_{2}^{sad3,4}\left( {{{\bar{h}}}_{c1}} \right)=0.4600 $
(\protect\ref{Emin34}) at the critical field
${{\bar{h}}_{c1}}=0.96 $: $\bar{h}=0.9599$, orange dot-dashed line
(1); $\bar{h}={{\bar{h}}_{c1}}=0.96$, black solid line (2);
$\bar{h}=0.96001$, red long-dashed line (3); $\bar{h}=0.9602$,
blue short-dashed line (4). For SOI constants  and magnetic field
direction we used the values $\bar{\protect\alpha}=0.8$,
$\bar{\protect\beta}=0.4$, ${{\protect\varphi }_{h}}=3\protect\pi
/4$.} \label{fig4}
\end{figure}

The positive roots of the equation (see Eq.(\ref{E12}))
\begin{equation}
{{\epsilon }_{1,2}}\left( \tilde{k},f\right) =\frac{{{\hbar }^{2}}{{{\tilde{k%
}}}^{2}}}{2m}-\frac{{{\hbar }^{2}}\tilde{k}}{m}{{\lambda }_{1,2}}\left(
f\right) +{{E}_{0}}=E  \label{Econtour}
\end{equation}%
describe the isoenergetic contours $k=k_{\pm }^{\left( j\right) }\left(
E,f\right) $ corresponding to physical electron states in the \textbf{k}%
-space for given energy $E$
\begin{equation}
k_{\pm }^{\left( 1,2\right) }={{\lambda }^{\left( 1,2\right) }}\pm \sqrt{{{%
\xi }^{\left( 1,2\right) }}},  \label{kcontour}
\end{equation}%
\begin{equation}
{{\xi }^{\left( 1,2\right) }}={{\left( {{\lambda }^{\left( 1,2\right) }}%
\right) }^{2}}+\frac{2m(E-{{E}_{0}})}{{{\hbar }^{2}}}\geq 0.
\label{ksicontour}
\end{equation}%
If $E>{{E}_{0}}$, the roots $k_{+}^{\left( 1,2\right) }>0$ for any values of
$f$, while roots $k_{-}^{\left( 1,2\right) }<0$, i.e. there are two contours
belonging to different energy branches. For $E<{{E}_{0}}$ real roots of
equation (\ref{Econtour}) exist, if the inequality $2m({{E}_{0}}-E)/{\hbar }%
^{2}\leq {{\left( {{\lambda }^{\left( 1,2\right) }}\right) }^{2}}$
holds. Roots $k_{\pm }^{\left( 1,2\right) }$ take positive values
for the angles $f$ at which ${{\lambda }^{\left( 1,2\right) }}>0$.
This means that the wave vector crosses the isoenergetic contour
twice.

\begin{figure}
\centering
\includegraphics[width=0.4\textwidth]{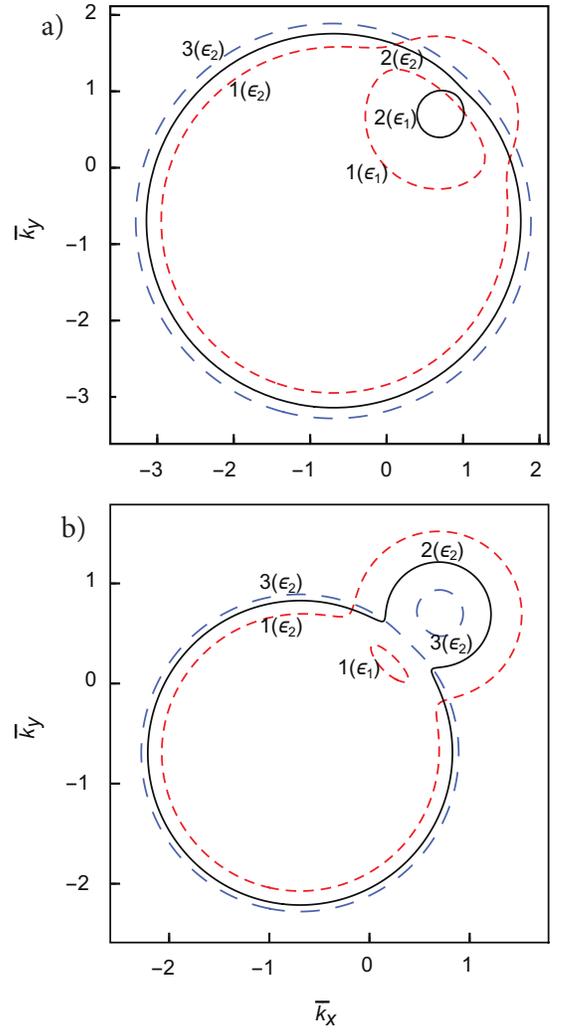}
\caption{a) Isoenergetic contours for both branches (labelled in
brackets) at the magnetic fields $\bar{h}\ge {{\bar{h}}_{c2}}=1$
and $\bar{E}=1$: $\bar{h}={{\bar{h}}_{c2}}$,
${{\bar{E}}_{0}}=0.5$, $\bar{E}>\bar{\protect\epsilon }_{1}^{\min
}=0.5 $, red short-dashed contours (1); $\bar{h}=1.45$,
${{\bar{E}}_{0}}=1.05$, $\bar{E}>\bar{\protect\epsilon }_{1}^{\min
}=0.95$, solid contours (2); $\bar{h}=1.8$,
${{\bar{E}}_{0}}=1.62$, $\bar{E}<\bar{\protect\epsilon }_{1}^{\min
}=1.3$, blue long-dashed contours (3). b) Isoenergetic contours
for ${{\bar{\protect\epsilon
}}_{1,2}}=\bar{E}={{\bar{E}}_{0}}\left( \bar{h}=0.5 \right)=0.125
$, $\bar{h}<{{\bar{h}}_{c1}}=0.96$, ${{\bar{h}}_{c2}}=1$:
$\bar{h}=0.3$, ${{\bar{E}}_{0}}=0.045<\bar{E}$, red short-dashed
contours (1); $\bar{h}=0.5$, ${{\bar{E}}_{0}}=0.125=\bar{E}$,
solid contour (2); $\bar{h}=0.6$, ${{\bar{E}}_{0}}=0.18>\bar{E}$,
blue long-dashed contours (3). For SOI constants and magnetic
field
direction we used the values $\bar{\protect\alpha}=0.8$,
$\bar{\protect\beta}=0.4$, ${{\protect\varphi }_{h}}=3\protect\pi /4$.} %
\label{fig5}
\end{figure}

\begin{figure}
\centering
\includegraphics[width=0.4\textwidth]{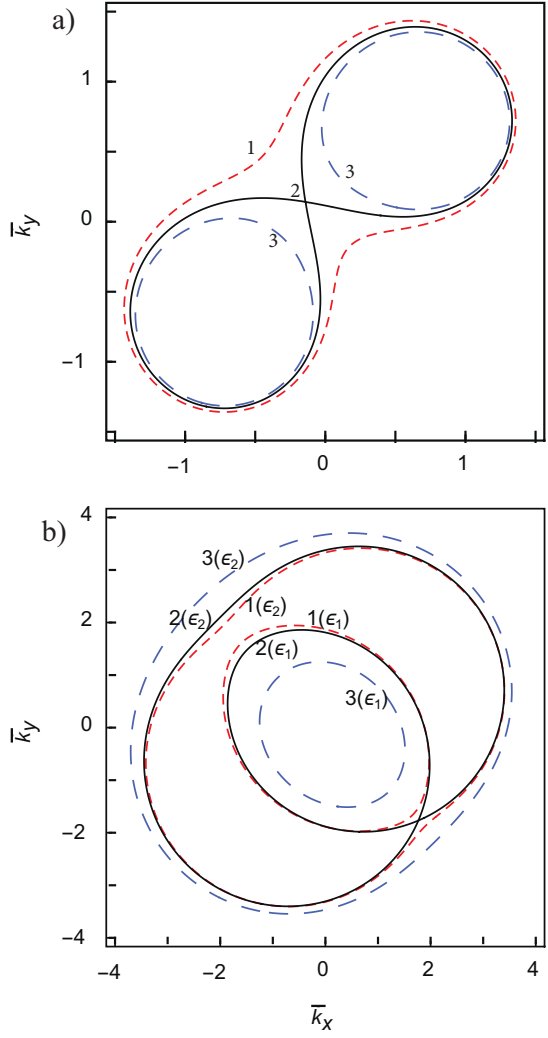}
\caption{a) Isoenergetic contours for the energy
$\bar{E}=\bar{\protect\epsilon }_{2}^{sad}\left( \bar{h}=0.3
\right)=0.28$ (\protect\ref{Esadllb}) and
${{\bar{h}}_{c2}}=0.04<\bar{h}<{{\bar{h}}_{c1}}=0.96$:
$\bar{h}=0.4$, red short-dashed contour(1); $\bar{h}=0.3$, black
solid contour (2); $\bar{h}=0.2$, blue long-dashed contours (3).
b) Isoenergetic contours for both branches (labelled in brackets)
at the energy $\bar{E}={{\bar{E}}_{0}}\left( \bar{h}=0.5
\right)=3.125$: $\bar{h}=0.1$, red short-dashed contours (1);
$\bar{h}=0.5$, black solid contour; $\bar{h}=2$, blue long-dashed
contours (3). For SOI constants  and magnetic field direction we
used the values $\bar{\protect\alpha}=0.8$,
$\bar{\protect\beta}=0.4$, ${{\protect\varphi }_{h}}=\protect\pi
/4$.} \label{fig6}
\end{figure}

\begin{figure}
\centering
\includegraphics[width=0.4\textwidth]{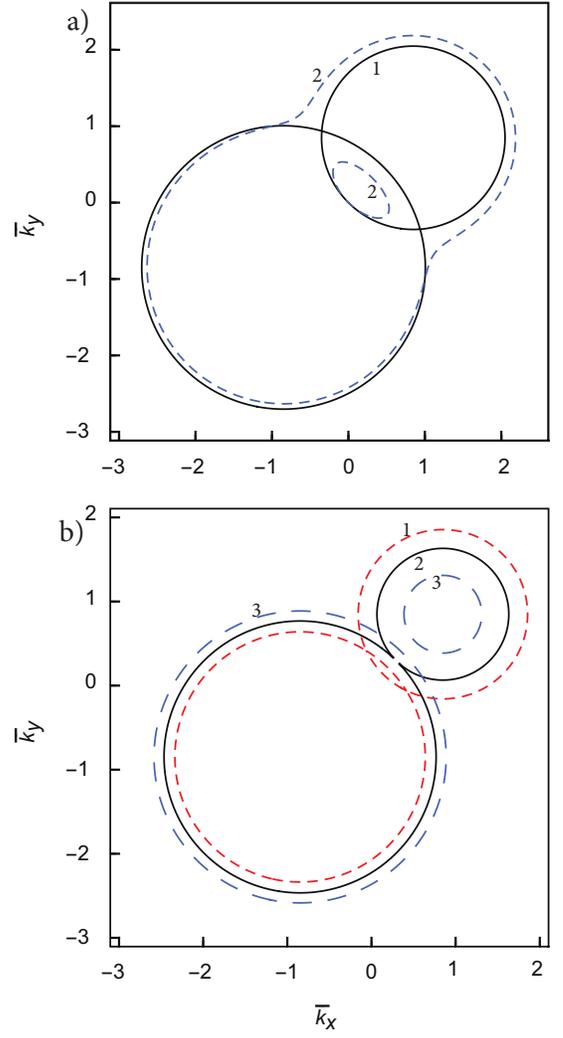}
\caption{a) Isoenergetic contours at $\protect\alpha
=\protect\beta $, $\overline{h}=0.5$, $\overline{E}=0.5$ for
different magnetic field directions: $\protect\varphi
_{h}=3\protect\pi /4$, solid contour (1); $\protect\varphi
_{h}=\protect\pi$, blue dashed contours (2). b) Isoenergetic
contours at and for the energy corresponding to the minimum of the
parabola (11) $\overline{\protect\epsilon }_{cont}^{\min
}=\overline{E}=0.0868$: $\overline{h}=0.3$, red short-dashed
contour (1); $\overline{h}=0.5$, solid contour (2);
$\overline{h}=0.7$, blue long-dashed contours (3). For SOI
constants we used the value $\overline{\protect\alpha
}=\overline{\protect\beta }=0.6$.} \label{fig7}
\end{figure}

In accordance with Vieta's formulas the roots of Eq. (\ref{Econtour}) obey
the relations
\begin{eqnarray}
&&k_{+}^{\left( 1,2\right) }k_{-}^{\left( 1,2\right) }=\frac{2m({{E}_{0}}-E)%
}{{{\hbar }^{2}}},  \notag \\
&&k_{+}^{\left( 1,2\right) }+k_{-}^{\left( 1,2\right) }=2{{\lambda }^{\left(
1,2\right) }},  \label{Viet}
\end{eqnarray}%
from which an interesting observation follows: at $E={{E}_{0}}$ the extremal
radii of the contours $k_{+}^{\left( 1,2\right) }=2{{\lambda }^{\left(
1,2\right) }}$, for which ${{\dot{\lambda}}^{\left( 1,2\right) }}=0$, give
energies (\ref{Ecrit}) and positions (\ref{kcr}) of critical points on total
surfaces $\epsilon ={{\epsilon }_{1,2}}$.

From the properties of energy spectrum which have been discussed in Sec. III
some general conclusions related to isoenergetic contours follow: 1) There
are no more when two separate contours for given energy $E$. 2) The contours
belonging to the branch $\epsilon ={{\epsilon }_{1}}$ exist for the energies
$E>{{E}_{0}}$ at $h<{{h}_{c2}}$ and $E>\epsilon _{1}^{\min }$ at $h>{{h}_{c2}%
}$. Contours $k=k_{\pm }^{\left( 1\right) }\left( f\right) $ are convex by
virtue of the equality (\ref{lambdaddot}). 3) In magnetic fields $h>{{h}_{c1}%
}$ the contour on the surface $\epsilon ={{\epsilon }_{2}}$ splits into two
separated contours for $E<\min \left( \epsilon _{2}^{sad1,2}\right) $.

At fixed energy $E$ magnetic field moves the energy of branch contact ${{E}%
_{0}}$ (\ref{E0}) and energies of critical points (\ref{Ecrit}) resulting in
Lifshitz electron transition of both types - appearance of new contour under
crossing the energy $E$ by minimum $\epsilon _{1,2}^{\min }$ and disruption
of the "neck" at $E=\epsilon _{2}^{sad}$.

At equal SOI constants $\alpha =\beta $ and magnetic field direction along
the axis ${{k}_{x}}=-{{k}_{y}}$ for energies $E>{{\hbar }^{2}}{{h}^{2}}/8m{{%
\alpha }^{2}}$ isoenergetic contours have two common contact points. Either
contour consists of two arcs of the radius (see Fig. 7),
\begin{equation}
{{k}^{\left( \pm \right) }}=\sqrt{\frac{2m}{{{\hbar }^{2}}}\left( E+\frac{2m{%
{\alpha }^{2}}}{{{\hbar }^{2}}}\mp h\right) }.  \label{ktochn}
\end{equation}%
Spin directions on each arc composing united contour are opposite, ${{\theta
}_{+}}=3\pi /4$ or ${{\theta }_{-}}=-\pi /4$, and arcs, which have the same
spin direction ${{\theta }_{\pm }}$ form total circumference.

Figures 4 and 5 illustrate some of the explicit results obtained in Sec. %
\ref{sec:level4}a for the magnetic field directed along the axis ${{k}_{x}}=-%
{{k}_{y}}$. Figure 4a shows the changes in the fine structure of
isoenergetic contours of the branch $\epsilon ={{\epsilon }_{2}}$ for the
energy close to the energy of saddle point $\epsilon _{2}^{sad3,4}\left(
h\right) $ (\ref{Emin34}) at the magnetic fields far from the critical
values $h<{{h}_{c1}},{h_{c2}}$ (\ref{f34h}). In this case we observe a
specific topological transition of the contour splitting in the transverse
to the "neck" direction. With the increase in the magnetic field the
electron and "hole" contours (curves 1) form unified critical contour with
two crossing points (curve 2). The critical contour breaks at the crossing
points in the transverse direction forming two electron contours (curves 3).
Figure 4b shows the disruption of the "neck" in the case when the energy $E<{%
{E}_{0}}\left( {{h}_{c1}}\right) $ equals the energy of the saddle point $%
E=\epsilon _{2}^{sad3,4}\left( {{h}_{c1}}\right) $ (\ref{Emin34}) at the
critical field ${{h}_{c1}}$ (\ref{f34h}). Two separate contours (curves 1)
touch at $h={{h}_{c1}}$ and in a higher field form a single non-convex
contour (curves 3,4).

Figure 5 illustrates another type of the topological transition in the
magnetic field: disappearance (or appearance) of a new detached region.
Isoenergetic contours for both branches at the magnetic fields $h\geq {{h}%
_{c2}}$ are shown. With an increase in the magnetic field the minimum $%
\epsilon _{1}^{\min }\left( h\right) $ of the branch $\epsilon ={{\epsilon }%
_{1}}$ moves up (curves 1 and 2) and at $E=\epsilon _{1}^{\min }\left(
h\right) $ crosses the energy level. At this field the contour related to
the branch $\epsilon ={{\epsilon }_{1}}$ disappears (curve 3). In Fig. 6b we
show the similar topological transition at $h<{{h}_{c2}}$ when ${{E}_{0}}$
is the smallest value of the branch $\epsilon ={{\epsilon }_{1}}$. If ${{E}%
_{0}}<E$ the energy spectrum consists of two electron contours (curves 1)
one of which disappears at ${{E}_{0}}=E$ (curve 2). In larger fields the
second contour splits into two contours (curves 3) at it was shown in Fig.
5a.

Figure 6 demonstrates the evolution of isoenergetic contour for the magnetic
field directed along the axis ${{k}_{x}}={{k}_{y}}$ (Sec. \ref{sec:level4}%
b). The disruption of the "neck" of the contour for the energy close to the
saddle point $\epsilon _{2}^{sad}\left( h\right) $ (\ref{Esadllb}) is shown
in Fig. 6a: The contour 1 corresponds to $\epsilon _{2}^{sad}\left( h\right)
<E$. The critical (self-crossing) contour 2 is in keeping with $\epsilon
_{2}^{sad}\left( h\right) =E$. In higher fields the critical contour splits
up into two disconnected parts (contours 3). Figure 6b demonstrates the
appearance of self-crossing contour in the magnetic field at which ${{E}_{0}}%
\left( h\right) =E$.

In Fig. 7a, we have shown a possibility of specific changes in the topology
by means of an in-plain rotation of the magnetic field in the case of equal
SOI constants: the self-crossing contour (2) splits into two split-off
contours (1,3) under a deflection of the magnetic field direction from
symmetry axis ${{k}_{x}}=-{{k}_{y}}$. Figure 7b shows the splitting of
self-crossing contour by the magnetic field for the energy close to minimal
energy of branch contact points $\epsilon _{cont}^{\min }\left( 0,h\right) ={%
{\hbar }^{2}}{{h}^{2}}/8m{{\alpha }^{2}}$ (\ref{Econt}).

\section{\label{sec:level6} Singularities in the electron density of states}

Density of states (DOS) singularities are related to the critical points of
energy spectrum. At the weak magnetic fields $\left( h<{{h}_{c1}},{{h}_{c2}}%
\right) $ results (\ref{Emin}) and (\ref{Esad}) obtained for 2D electrons
with R-D SOI show that energies of both minima and saddle points move in
opposite directions on the energy scale with the increasing of the value $h$%
. So, the number of van Hove's singularities is doubled by the magnetic
field. Exclusions are the directions of vector $\mathbf{h}$ along symmetry
axes when two minima (\ref{Emin12b}) $\left( {{\varphi }_{h}}=\pi /4,-3\pi
/4\right) $ or two saddle points (\ref{Esad}) $\left( {{\varphi }_{h}}=3\pi
/4,-\pi /4\right) $ "synchronously" move with a change in the magnetic
field. In these cases the DOS has three singular points. At $h>{{h}_{c1}}$
the DOS contains two singularities which are associated with two minimum
points at $h>{{h}_{c2}}$ or minimum and saddle points at $h<{{h}_{c2}}$.

By using the coordinates (\ref{ktilde}) the DOS can be found from the
relations \cite{KozKolesn2018}
\begin{equation}
\rho \left( E\right) =\frac{m}{\pi {{\hbar }^{2}}};\quad E\geq {{E}_{0}},
\label{rhoconst}
\end{equation}%
\begin{equation}
\rho \left( E\right) =\frac{m}{2\pi ^{2}{{\hbar }^{2}}}\sum_{j=1,2}\oint df%
\frac{\lambda ^{\left( j\right) }}{\sqrt{\xi ^{\left( j\right) }}}\Theta
\left( \lambda ^{\left( j\right) }\right) \Theta \left( \xi ^{\left(
j\right) }\right) ;E\leq E_{0},  \label{rhrogeneral}
\end{equation}%
where ${{\lambda }^{\left( j\right) }}$ and ${{\xi }^{\left( j\right) }}$
are defined by Eqs. (\ref{lambda}) and (\ref{ksicontour}), $\Theta \left(
x\right) $ is Heaviside step function. Equation (\ref{rhoconst}) shows that
the DOS is the same as for free 2D electron gas for energies $E\geq {{E}_{0}}
$. When the opposite inequality $E<{{E}_{0}}$ the DOS $\rho \left( E\right) $
depends on the magnetic field and constants of SOI.

The electron density one find by integration of
Eq.(\ref{rhrogeneral}) over energies below Fermi level ${{E}_{F}}$
\cite{KozKolesn2018}
\begin{equation}
{{n}_{e}}=\frac{m}{\pi {{\hbar }^{2}}}\left[
{{E}_{F}}+\frac{m}{2{{\hbar }^{2}}}\left( {{\alpha }^{2}}+{{\beta
}^{2}} \right) \right] \label{ne}
\end{equation}
for $ {{E}_{F}}\ge {{E}_{0}} $.

The Van Hove singularities of $\rho \left( E\right) $ are related
to minima and saddle points on the energy surfaces
(\ref{Econtour}) and correspond to
singularities in the integral (\ref{rhrogeneral}) ${{\xi }^{\left( j\right) }%
}\left( f_{\nu }^{\left( j\right) }\right) =0$. In order to separate the
singular part of DOS near the critical point $E\rightarrow \epsilon
_{i}^{crit},\ f\rightarrow f_{\nu }^{\left( i\right) }$ we use a standard
way. Equation (\ref{rhrogeneral}) can be written as a sum of a convergent
and divergent parts
\begin{equation}
\rho \left( E\right) ={{\rho }_{0}}\left( E\right) +\delta \rho \left(
E\right) ,  \label{rhofull}
\end{equation}%
\begin{eqnarray}
{{\rho }_{0}}\left( E\right) &=&\frac{m}{2{{\pi }^{2}}{{\hbar }^{2}}}%
\sum\limits_{j=1,2}{\oint {d}f}\frac{{{\lambda }^{\left( j\right) }}\left(
f\right) -{{\delta }_{ij}}{{\lambda }^{\left( j\right) }}\left( f_{\nu
}^{\left( j\right) }\right) }{\sqrt{{{\xi }^{\left( j\right) }}\left(
f\right) }}  \label{rho0} \\
&&\times \Theta \left( {{\lambda }^{\left( j\right) }}\right) \Theta \left( {%
{\xi }^{\left( j\right) }}\right) ,
\end{eqnarray}%
\begin{equation}
\delta \rho \left( E\right) =\frac{m}{2{{\pi }^{2}}{{\hbar }^{2}}}{{\lambda }%
^{\left( i\right) }}\left( f_{\nu }^{\left( i\right) }\right) \oint {\frac{df%
}{\sqrt{{{\xi }^{\left( i\right) }}\left( f\right) }}\Theta \left( {{\lambda
}^{\left( i\right) }}\right) \Theta \left( {{\xi }^{\left( i\right) }}%
\right) }.  \label{rhodelta}
\end{equation}%
The continuous function ${{\xi }^{\left( i\right) }}\left( f\right) $ can be
expanded as a Taylor series in a vicinity of the point $f=f_{\nu }^{\left(
i\right) }$
\begin{align}
& {{\xi }^{\left( 1,2\right) }}\left( f\right) ={{\left( {{\lambda }^{\left(
1,2\right) }}\right) }^{2}}+\frac{2m(E-{{E}_{0}})}{{{\hbar }^{2}}}\simeq
\frac{2m(E-\epsilon _{1,2}^{crit})}{{{\hbar }^{2}}}  \notag \\
& +{{\lambda }^{\left( 1,2\right) }}\left( f_{\nu }^{\left( 1,2\right)
}\right) {{{\ddot{\lambda}}}^{\left( 1,2\right) }}\left( f_{\nu }^{\left(
1,2\right) }\right) {{\left( f-f_{\nu }^{\left( 1,2\right) }\right) }^{2}}
\notag \\
& +\frac{1}{6}{{\lambda }^{\left( 1,2\right) }}\left( f_{\nu }^{\left(
1,2\right) }\right) {{}^{\left( 1,2\right) }}\left( f_{\nu }^{\left(
1,2\right) }\right) {{\left( f-f_{\nu }^{\left( 1,2\right) }\right) }^{3}}%
+...\ .  \label{ksiseries}
\end{align}%
Substituting the expansion (\ref{ksiseries}) into Eq. (\ref{rhodelta}) we
integrate using the cutoff of the integral by $\Theta $-functions. At $%
E\rightarrow \epsilon _{i}^{crit}$ the singular term $\delta \rho \left(
E\right) $ (\ref{rhodelta}) does not depend on the interval of integration.
As shown in Sec. \ref{sec:level3} for the energy minima $\epsilon
_{1,2}^{\min }$ the second derivatives are negative, ${{\ddot{\lambda}}^{\left( 2\right) }}%
\left( f_{\nu }^{\left( 1,2\right) }\right) <0$, and for the saddle points $%
\epsilon _{2}^{sad}$ the second derivatives are positive
${{\ddot{\lambda}}^{\left( 2\right) }}\left( f_{\nu }^{\left(
1,2\right) }\right) >0$. As a result we have in a vicinity of the
minimum points
\begin{equation}
\delta \rho \left( E\right) =\frac{m}{2\pi {{\hbar }^{2}}}{{\left. \sqrt{%
\frac{{{\lambda }^{\left( 1,2\right) }}}{\left\vert {{{\ddot{\lambda}}}%
^{\left( 1,2\right) }}\right\vert }}\right\vert }_{f=f_{\nu }^{\left(
1,2\right) }}}\Theta \left( E-\epsilon _{1,2}^{\min }\right) .
\label{rhotheta}
\end{equation}%
For saddle point one finds
\begin{eqnarray}
\delta \rho \left( E\right) &=&-\frac{m}{2{{\pi }^{2}}{{\hbar }^{2}}}\sqrt{%
\frac{{{\lambda }^{\left( 2\right) }}}{{{{\ddot{\lambda}}}^{\left( 2\right) }%
}}}  \notag \\
&&\times \ln {{\left[ \frac{2m\left\vert E-\epsilon _{2}^{sad}\right\vert }{{%
{\hbar }^{2}}{{\lambda }^{\left( 2\right) }}{{{\ddot{\lambda}}}^{\left(
2\right) }}}\right] }_{f=f_{\nu }^{\left( 2\right) }}}.  \label{rholn}
\end{eqnarray}%
At the critical magnetic field $h={{h}_{c1}}$ the first and second
derivatives of ${{\lambda }^{\left( 2\right) }}\left( f\right) $ are equal
to zero, and the third term in the expansion (\ref{ksiseries}) must be taken
into account. The singular part of the DOS in the case of the degenerate
critical point reads ($\dddot{\lambda}^{(2)}\neq 0$)
\begin{eqnarray}
\delta \rho \left( E\right) &=&\frac{m\sqrt{6}}{2{{\pi }^{3/2}}{{\hbar }^{2}}%
}\frac{\Gamma \left( {\scriptstyle{}^{7}/{}_{6}}\right) }{\Gamma \left( {%
\scriptstyle{}^{2}/{}_{3}}\right) }  \notag \\
&&\times \sqrt{\frac{{{\lambda }^{\left( 2\right) }}}{\left\vert {{}\dddot{%
\lambda}^{\left( 2\right) }}\right\vert }}\left( \frac{12m\left\vert
E-\epsilon _{2}^{crit}\right\vert }{{{\hbar }^{2}}{{\lambda }^{\left(
2\right) }}\left\vert \dddot{\lambda}{{}^{\left( 2\right) }}\right\vert }%
\right) _{f=f_{\nu }^{\left( 2\right) }}^{-1/6}.  \label{rhoc1}
\end{eqnarray}

The results (\ref{rhotheta}),(\ref{rholn}) agree with the classical results
for the two-dimensional case obtained in van Hove's paper \cite{vanHove1953}%
. The simple relations between the energies of critical points and SOI
constants for directions of magnetic field along symmetry axes (see Sec.IV)
is the way to find $\alpha $ and $\beta $ from the position of DOS
singularities on the magnetic field scale.

\section{\label{sec:level7} Conclusions}

The evolution of energy spectrum of 2D electron gas with combined Rashba and
Dresselhaus SOI (\ref{eigenvalues}) under the influence of in-plain magnetic
field $\mathbf{B}$ has been analyzed for arbitrary SOI constants. It has
been shown that geometry of energy surfaces (\ref{E12}) and isoenergetic
contours (\ref{Econtour}) can be described by means of a single function (%
\ref{lambda}), which depends on the magnetic field and SOI constants. We
have found the relations which describe dependencies of critical point
energies (\ref{Ecrit}) and their positions (\ref{kcr}) in the wave vector
space on the vector $\mathbf{B}$. There are two critical values of magnetic
field at which the essential transformation of the energy spectrum occurs:
At the field $B={{B}_{1}}$ (\ref{lambdaziro}) the minimum point and saddle
point of the energy branch $\epsilon ={{\epsilon }_{2}}$ "annihilate" and at
$B={{B}_{2}}$ (\ref{hc2}) the conical point of the branch $\epsilon ={{%
\epsilon }_{1}}$ transforms into the critical (minimum) point. Finally, the
spectrum having four critical points (two minima and two saddle points) and
a conical point at $B=0$ evolves into spectrum with two minima at $B>{{B}_{1}%
},{{B}_{2}}$. The general conclusions are illustrated for the directions of
vector $\mathbf{B}$ along the symmetry axes. On the basis of an analysis of
spectrum critical points dependence on the magnetic field Lifshitz
topological transitions in the geometry of isoenergetic contours have been
studied (Figs.4-7). Along with critical contours related to spectrum
critical points the appearance (or disappearance) of self - crossing
contours with a magnetic field variation is found as well. Singular
additions to the electron density of states have been derived (\ref{rhotheta}%
) - (\ref{rhoc1}). The positions of these singularities on the magnetic
field scale make possible to find both the SOI constant. We have found $%
\left( -1/6\right) $-root singularity for the degenerate critical points at $%
B={{B}_{c1}}$ (\ref{rhoc1}). The obtained results can be used for
theoretical investigations of any kinetic and thermodynamic characteristics
of 2D electrons with R-D SOI in the in-plane magnetic field as well as for
interpretation of experimental data.

Magnetic-field-driven topological transitions can be observed in
the 2D electron gas with a low electron density $n\simeq
10^{9}\div 10^{10}cm^{-2}$ (see, for example, \cite{Zhu}). In
heterostructures with higher density it could be essentially
reduced by a negative gate voltage \cite{Rossler}. For the typical
values of R-D SOI constants and an effective mass for
Al$_{x}$Ga$_{1-x}$N/GaN heterostructure, $\alpha \simeq
10^{-10}eV\cdot cm,$ $\alpha
/\beta \simeq 10$, $m=0.2m_{0}$, $g^{\ast }=2$ \ \cite{Chunming,Knap} and $%
n\simeq 10^{10}cm^{-2}$, we estimate a Fermi energy, $E_{F}\simeq
0.1meV$, by using Eq. (\ref{ne}). In this case the van Hove
singularities appear in magnetic field $0<B\lesssim 2T.$ Other
possibility to observe the predicted topological transitions is \
the in-plane tunnelling spectroscopy
\cite{Yee,Pairor2008,Pairor2013}. While a tunnelling conductances
is proportional to the density of states at a shifted energy $\rho
\left( \epsilon =E_{F}-eU\right) $, where $eU$ is a bias energy,
the electron states below Fermi level can be investigated.

\begin{acknowledgments}
One of us (Yu.K) would like to acknowledge useful discussion with S. V.
Kuplevakhsky.
\end{acknowledgments}

\appendix*
\section{Functions ${{\lambda }^{\left( 1,2\right) }}$ and ${{\ddot{\lambda}}^{\left( 1,2\right)
}} $ in the critical points of energy spectrum ${{\dot{\lambda}
}^{\left( 1,2\right) }}=0$ for the magnetic field directed along
symmetry axis.}

\setcounter{equation}{0}
\renewcommand\theequation{A.\arabic{equation}}

\textsl{Magnetic field directed along ${{k}_{x}}=-{{k}_{y}}$ axis
}

Substituting the angles (\ref{fsim}),(\ref{f34}) corresponding to
the zeros of $\dot{\lambda}{^{\left( 1,2\right) }} $ we find the
functions ${{\lambda }^{\left( 1,2\right) }}$ (\ref{lambda})
\begin{equation}
{{\lambda }^{\left( 1,2\right) }}\left( -\frac{3\pi }{4}\right) =\mp \frac{m%
}{{{\hbar }^{2}}}\left( \alpha +\beta \right) \mp \frac{h}{\alpha
+\beta }, \label{fsimlambda}
\end{equation}%
\begin{equation}
{{\lambda }^{\left( 1,2\right) }}\left( \frac{\pi }{4}\right) =\mp \frac{m}{{%
{\hbar }^{2}}}\left( \alpha +\beta \right) \pm \frac{h}{\alpha
+\beta }, \label{laPi4}
\end{equation}%
\begin{equation}
{{\lambda }^{\left( 1,2\right) }}\left( {{f}_{3,4}}\right) =\mp \frac{m}{{{h}%
^{2}}}(\alpha -\beta
)\sqrt{1-\frac{{{h}^{2}}}{{{h}_{c1}}{{h}_{c2}}}},\quad h\leq
{{h}_{c1}},  \label{f34lambda}
\end{equation}%
and second derivatives ${{\ddot{\lambda}}^{\left( 1,2\right)
}}\left(
f\right) $%
\begin{equation}
{{\ddot{\lambda}}^{\left( 1,2\right) }}\left( -\frac{3\pi
}{4}\right) =\pm \frac{h+{{h}_{c1}}}{\alpha +\beta },
\label{fsimdot}
\end{equation}%
\begin{equation}
{{\ddot{\lambda}}^{\left( 1,2\right) }}\left( \frac{\pi
}{4}\right) =\mp \frac{h-{{h}_{c1}}}{\alpha +\beta },
\label{ladot2}
\end{equation}%
\begin{gather}
{{\ddot{\lambda}}^{\left( 1,2\right) }}\left( {{f}_{3,4}}\right) =\mp \frac{{%
{h}_{c1}}}{\alpha -\beta }\left( 1-{{\left( \frac{h}{{{h}_{c1}}}\right) }^{2}%
}\right)  \notag \\
\times \sqrt{1-\frac{{{h}^{2}}}{{{h}_{c1}}{{h}_{c2}}}};\quad h\leq {{h}_{c1}}%
.  \label{f34dot}
\end{gather}%
The Eqs. (\ref{Ecrit}), (\ref{fsimlambda}) - (\ref{f34lambda})
give the
formulas for energies which can be possible critical points ${{\epsilon }%
^{crit\ \nu }}\left( h\right) ={{\epsilon }_{1,2}}\left( f_{\nu
}^{\left( 1,2\right) }\right) $ (\ref{Ecrit}) of energy spectrum
\begin{eqnarray}
&&\epsilon ^{crit1}\left( h\right) =-\frac{m}{2{{\hbar
}^{2}}}{{\left(
\alpha +\beta \right) }^{2}}-h,  \notag \\
&&\epsilon ^{crit2}\left( h\right) =-\frac{m}{2{{\hbar
}^{2}}}{{\left( \alpha +\beta \right) }^{2}}+h,  \label{Ecrit12}
\end{eqnarray}%
\begin{equation}
{{\epsilon }^{crit3,4}}\left( h\right) =-\frac{m}{2{{\hbar
}^{2}}}{{\left( \alpha -\beta \right)
}^{2}}+\frac{{{h}^{2}}{{\hbar }^{2}}}{8m\alpha \beta }.
\label{Ecrit34}
\end{equation}%

\textsl{Magnetic field directed along ${{k}_{x}}={{k}_{y}}$ axis.
}

The functions ${{\lambda }^{\left( 1,2\right) }}$ in the points (\ref{f1234b}%
) of first derivatives ${{\dot{\lambda}}^{\left( 1,2\right) }}$
zeros are
given by%
\begin{equation}
{{\lambda }^{\left( 1,2\right) }}\left( \frac{3\pi }{4}\right) =\mp \frac{m}{%
{{\hbar }^{2}}}\left( \alpha -\beta \right) +\frac{h}{\alpha
-\beta }, \label{lambda12b}
\end{equation}%
\begin{equation}
{{\lambda }^{\left( 1,2\right) }}\left( -\frac{\pi }{4}\right) =\mp \frac{m}{%
{{\hbar }^{2}}}\left( \alpha -\beta \right) -\frac{h}{\alpha
-\beta }, \label{lambda12c}
\end{equation}%
\begin{equation}
{{\lambda }^{\left( 1,2\right) }}\left( {{f}_{3,4}}\right) =\mp \frac{m}{{{h}%
^{2}}}(\alpha +\beta
)\sqrt{1+\frac{{{h}^{2}}}{{{h}_{c1}}{{h}_{c2}}}},\quad h\leq
{{h}_{c1}},  \label{lambda34b}
\end{equation}%
and their second derivatives ${{\ddot{\lambda}}^{\left( 1,2\right) }}$ read%
\begin{equation}
{{\ddot{\lambda}}^{\left( 1,2\right) }}\left( \frac{3\pi }{4}\right) =-\frac{%
h-{{h}_{c1}}}{\alpha -\beta },  \label{lambda12dotb}
\end{equation}%
\begin{equation}
{{\ddot{\lambda}}^{\left( 1,2\right) }}\left( -\frac{\pi }{4}\right) =\frac{%
h+{{h}_{c1}}}{\alpha -\beta },  \label{lambda12dotc}
\end{equation}%
\begin{eqnarray}
&&{{\ddot{\lambda}}^{\left( 1,2\right) }}\left( {{f}_{3,4}}\right)
=\pm
\frac{{{h}_{c1}}}{\alpha +\beta }\left( 1-{{\left( \frac{h}{{{h}_{c1}}}%
\right) }^{2}}\right)  \notag \\
&&\times \sqrt{1+\frac{{{h}^{2}}}{{{h}_{c1}}{{h}_{c2}}}};\quad h\leq {{h}%
_{c1}}.  \label{lambda34dotb}
\end{eqnarray}%
As the last step one must separate the physical solutions: Only
the positive values of ${{\lambda }^{\left( 1,2\right) }}\left(
f_{\nu }^{\left( 1,2\right) }\right) $ satisfy to the Eq.
(\ref{dEdk}). From the simple analysis of Eqs.
(\ref{fsimlambda})-(\ref{lambda34dotb}) we obtain conditions of
the extrema existence in different ranges of magnetic field.


\begin{thebibliography}{99}
\bibitem{Lifshitz1960} I.M. Lifshitz, Sov. Phys. JETP, \textbf{11}, 1130
(1960).

\bibitem{LAK} I.M. Lifshits, M.Ya. Azbel', and M.I. Kaganov, Electron
Theory of Metals (Consultants Bureau, New York), (1973).

\bibitem{Varlamov1989} A.A. Varlamov, V.S. Egorov, and A.V. Pantsulaya,
Adv. Phys. \textbf{38}, 469 (1989).

\bibitem{Blanter1994} Ya.M. Blanter, M.I. Kaganov, A.V. Pantsulaya, A.A.
Varlamov, Physics Reports \textbf{245} , 159-257 (1994).

\bibitem{Volovik2018} G.E. Volovik, Low Temp. Phys. \textbf{43}, 47 (2017);
UFN, \textbf{188}, 95 (2018).

\bibitem{Manchon2015} A. Manchon, H. C. Koo, J. Nitta, S. M. Frolov and R.
A. Duine, Nature Materials, \textbf{14}, 871 (2015).

\bibitem{Bercioux2015} D. Bercioux, P. Lucignano, Report on Progress in
Physics, \textbf{78} 106001 (2015).

\bibitem{Boiko} I.I. Boiko, E.I. Rashba. Sov. Phys. Solid State \textbf{2(8)}%
, 1692 (1961).

\bibitem{Schober} G.A.H. Schober, H. Murakawa, M.S. Bahramy, R. Arita, Y.
Kaneko, Y. Tokura, and N. Nagaosa, Phys. Rev. Lett. \textbf{108}, 247208
(2012).

\bibitem{Meier2007} L. Meier, G. Salis, I. Shorubalko, E. Gini, S. Scho and
K. Ensslin, Nature Physics, \textbf{3}, 650 (2007).

\bibitem{Hao2012} Y.F. Hao, Eur. Phys. J. B 85: 84 (2012).

\bibitem{Silveira2016} L. Silveira, P. Barone, and S. Picozzi, Phys. Rev. B
\textbf{93}, 245159 (2016).

\bibitem{Kepenekian2017} M. Kepenekian and J. Even, J. Phys. Chem. Lett.,
\textbf{8} (14), 3362 (2017).

\bibitem{Ganichev2014} S. D. Ganichev, and L. E. Golub, Phys. Status Solidi
B \textbf{251}, 1801 (2014).

\bibitem{Wang2010} C.M. Wang Phys. Rev. B \textbf{82}, 165331 (2010).

\bibitem{Wilde}M.A. Wilde and D. Grundler, New Journal of Physics \textbf{15} 115013
(2013).

\bibitem{BermanFlatte2010} D.H. Berman, M.E. Flatte, Phys. Rev. Lett.
\textbf{105}, 157202 (2010).

\bibitem{Badalyan2010} S.M. Badalyan, A. Matos-Abiague, G. Vignale, and J.
Fabian, Phys. Rev. B \textbf{81}, 205314 (2010).

\bibitem{KozKolesnLTP} I.V. Kozlov, Yu.A. Kolesnichenko, Low Temp. Phys.
\textbf{43}, 855 (2017).

\bibitem{Winkler2003} R. Winkler, Spin-Orbit Coupling Effects in
Two-Dimensional Electron and Hole Systems (Springer-Verlag, Berlin
Heidelberg, 2003).

\bibitem{Pershin2004} Yu.V. Pershin, J.A. Nesteroff, and V. Privman, Phys.
Rev. B \textbf{69}, 121306(R) (2004).

\bibitem{Tkach2016} Yu.Ya. Tkach, JETP Letters \textbf{104,} 105 (2016).

\bibitem{ShevchenkoKopeliovich2016} O.N. Shevchenko and A.I. Kopeliovich,
Low Temp. Phys., \textbf{42}, 196 (2016).

\bibitem{Alekseev2007} P.S. Alekseev, M.V. Yakunin and I. N. Yassievich,
Semiconductors, \textbf{41}, 1092 (2007).

\bibitem{Sablikov} V. A. Sablikov and Yu. Ya. Tkach, Phys. Rev. B, \textbf{99}, 035436 (2019).

\bibitem{vanHove1953} L. van Hove, Phys. Rev. \textbf{89}, 1189 (1953).

\bibitem{Wolff} K. Wolff,  R. Schafer, M. Meffert, D. Gerthsen, R. Schneider, and D. Fuchs,
Phys. Rev. B, \textbf{95}, 245132 (2017).

\bibitem{Eppenga} R. Eppenga and M. F. H. Schuurmans,Phys. Rev. B, \textbf{37%
},10923 (1988).

\bibitem{Gilbertson} A.M. Gilbertson, M. Fearn, J.H. Jefferson, B.N.
Murdin, P.D. Buckle, and L.F. Cohen, Phys. Rev. B, \textbf{77},
165335 (2008).

\bibitem{Kleinert} P. Kleinert, V.V. Bryksin, Phys. Rev. B, \textbf{76},
205326 (2007).

\bibitem{Wang} C.M. Wang, Phys. Rev. B, \textbf{82} 165331 (2010).

\bibitem{Wen} Wen Xu , Yong Guo, Physics Letters A, \textbf{340} 281(2005).

\bibitem{Rashba1984} E.I. Rashba, Sov. Phys. Solid State \textbf{2}, 1109
(1960), Yu. A. Bychkov, and E. I. Rashba, JETP Lett. \textbf{39}, 78 (1984).

\bibitem{Dresselhaus1955} G. Dresselhaus, Phys. Rev. \textbf{100}, 580 (1955).

\bibitem{Nitta1997} Junsaku Nitta, Tatsushi Akazaki, and Hideaki Takayanagi,
Phys. Rev. Lett. \textbf{78}, 1335 (1997).

\bibitem{Sato2001} Y. Sato, T. Kita, S. Gozu, and S. Yamada, J. Appl. Phys.
\textbf{89}, 8017 (2001).

\bibitem{Beukman2017} A.J.A. Beukman, F.K. de Vries, J. van Veen, etc.,
Phys. Rev. B \textbf{96}, 241401 (2017).

\bibitem{KozKolesn2018} I.V. Kozlov, Yu.A. Kolesnichenko, arxiv.org.
1805.05699v1 (2018); Physics Letters A, \textbf{383}, 764 (2019).

\bibitem{Zhu}J. Zhu, H.L. Stormer, L.N. Pfeiffer, K.W. Baldwin,
and K.W.West, Phys. Rev. B. \textbf{90}, 056805 (2003).

\bibitem{Rossler}C. Rossler, T. Feil, P. Mensch, T. Ihn, K. Ensslin, D Schuh,
and W. Wegscheider, New Journal of Physics \textbf{12 }043007
(2010).

\bibitem{Chunming}Chunming Yin, Bo Shen, Qi Zhang, Fujun Xu, Ning Tang, Longbin Cen,
Xinqiang Wang, Yonghai Chen, and Jinling Yu, Applied Phys. Lett.
\textbf{97}, 181904 (2010).

\bibitem{Knap}W. Knap, E. Frayssinet, and M.L. Sadowski, C. Skierbiszewski,
D. Maude, V. Falko, M. Asif Khan, M. Asif Khan, Applied Phys.
Lett. \textbf{75}, 3156 (1999).

\bibitem{Yee}Yee Sin Ang, Zhongshui Ma and C. Zhang, Scientific Reports \textbf{4}, 3780
(2014).

\bibitem{Pairor2008}B. Srisongmuang, P. Pairor, Phys. Rev. B. \textbf{78}, 155317 (2008).

\bibitem{Pairor2013}A. Jantayod, P.Pairor, Physica E, \textbf{48}, 111 (2013).

\end{thebibliography}
\end{document}